\documentclass{article}

\usepackage{natbib}
 \bibpunct[, ]{(}{)}{,}{a}{}{,}%

\usepackage{arxiv}

\usepackage{caption}
\usepackage{subcaption}

\usepackage{nameref}
\usepackage{booktabs}
\usepackage{appendix}
\usepackage{array}
\usepackage{float}
\usepackage{amsfonts}
\usepackage{nicefrac}       
\usepackage{microtype}      
\usepackage{graphicx}
\usepackage{doi}
\usepackage{mathtools}

\usepackage[utf8]{inputenc} 
\usepackage[T1]{fontenc}    
\usepackage{hyperref}       
\usepackage{url}            

\usepackage{setspace}
\usepackage{appendix}

\usepackage[export]{adjustbox}

\renewcommand{\headeright}{}
\renewcommand{\undertitle}{Forthcoming in Management Science}

\mathtoolsset{showonlyrefs=true}

\date{September 16, 2022}

\begin{document}



\title{Sequential Learning and Economic Benefits from Dynamic Term Structure Models}




\author{Tomasz Dubiel-Teleszynski \\
	Department of Statistics, London School of Economics\\
	Houghton Street, London, WC2A 2AE, United Kingdom \\
	\texttt{t.dubiel-teleszynski1@lse.ac.uk} \\
	\And
	Konstantinos Kalogeropoulos \\
	Department of Statistics, London School of Economics\\
	Houghton Street, London, WC2A 2AE, United Kingdom \\
	\texttt{k.kalogeropoulos@lse.ac.uk} \\
	\And
	Nikolaos Karouzakis \\
	ALBA Graduate Business School, The American College of Greece\\
	6-8 Xenias Str, 115 28 Athens, Greece \\
	\& University of Sussex Business School, Brighton, United Kingdom \\
	\texttt{nkarouzakis@alba.acg.edu} \\
}

\maketitle

\begin{abstract}
We explore the statistical and economic importance of restrictions on the dynamics of risk compensation from the perspective of a real-time Bayesian learner who predicts bond excess returns using dynamic term structure models (DTSMs). The question on whether potential statistical predictability offered by such models can generate economically significant portfolio benefits out-of-sample, is revisited while imposing restrictions on their risk premia parameters. To address this question, we propose a methodological framework that successfully handles sequential model search and parameter estimation over the restriction space in real time, allowing investors to revise their beliefs when new information arrives, thus informing their asset allocation and maximising their expected utility. Empirical results reinforce the argument of sparsity in the market price of risk specification since we find strong evidence of out-of-sample predictability only for those models that allow for level risk to be priced and, additionally, only one or two of these risk premia parameters to be different than zero. Most importantly, such statistical evidence is turned into economically significant utility gains, across prediction horizons, different time periods and portfolio specifications. In addition to identifying successful DTSMs, the sequential version of the stochastic search variable selection (SSVS) scheme developed can be applied on its own and also offer useful  diagnostics monitoring key quantities over time. Connections with predictive regressions are also provided. 
\end{abstract}



\section{Introduction}
\label{Intro}


\subsection{Restrictions and out-of-sample Economic Benefits}

Accurately estimating and forecasting bond risk premia, in real time, is of central economic importance for the transmission mechanism of monetary policy as well as for investors' portfolio strategies. Even more important is understanding and identifying the contribution of risk premia to longer term interest rates, which largely depends on our ability to accurately infer expectations for the future path of the short end of the yield curve\footnote{See, \cite{Kim05} and \cite{Cochrane09}, among others, for studies that attempt to decompose forward rates into expectations of short rates and risk premia.}. To successfully do so, it is essential to account for no-arbitrage, which implies restrictions on the cross-sectional and time series dynamics of the term structure (see, \cite{Joslin11} and \cite{Bauer18}). The latter are largely exploited in related literature by dynamic term structure models (DTSMs), which impose tight restrictions on the dynamics of risk compensation, an essential component of the models. Failure to impose such restrictions, as in the unrestricted maximally flexible model widely used by almost all existing studies, leads to absence of no-arbitrage and, as such, to the generation of artificially stable short rate expectations and highly volatile risk premia \citep{KimOrphanides12,Bauer18}. 

The importance of the market price of risk specification, and the associated restrictions related to it, has been extensively studied in earlier research (see, \cite{Dai00}, \cite{Duffee02}, \cite{Ang03}, \cite{Kim05}), which has mainly focused on imposing ad hoc\footnote{Ad-hoc restrictions, are used in \cite{Dewachter06}, and \cite{Rudebusch08}, among others. Furthermore, the route of imposing prior restrictions is followed by \cite{Ang07}.} zero restrictions on the parameters governing the dynamics of the risk premia\footnote{A common practice used is to, first, estimate an unrestricted maximally flexible model, and at a second step, to re-estimate it by setting to zero those parameters that have large standard errors. According to \cite{Bauer18}, such an approach often leads to the wrong model.}. 
This practice, however, has been criticised (see, \cite{Kim12} and \cite{Bauer18}), since it raised concerns about, first, the joint significance of the constraints, second, the magnitude of the associated standard errors\footnote{According to \cite{Kim12}, it is unclear how small these have to be in order to set a parameter to zero.} and, third, the failure to provide meaningful economic justification for the estimated parameters and the resulting state variables.
Only recently, a few studies have investigated more systematic approaches to imposing restrictions on the dynamics of risk compensation\footnote{See, \cite{Cochrane09} for a 4-factor affine model, \cite{Joslin14} for an unspanned macro-finance DTSM, and \cite{Duffee11} and \cite{Bauer18} for yields-only versions of DTSMs. An alternative approach is followed by \cite{Chib09}, who impose strong prior restrictions such that the yield curve is (on average) upward sloping, an assumption that is empirically and economically plausible.}. In particular, \cite{Cochrane09} and \cite{Duffee11} introduce tight restrictions, driven by prior empirical analysis, while \cite{Joslin14} select zero restrictions. \cite{Bauer18} promotes the use of Bayesian variable selection samplers to identify such restrictions on risk prices.

Although the literature has noted the importance of restrictions, yet, no study has, so far, addressed and quantified their statistical and economic importance, out-of-sample\footnote{Empirical tests in \cite{Duffee11} suggest that the choice of no-arbitrage restrictions does not influence the out-of-sample performance of the models, given that they produce forecasts with indistinguishable differences.}. Most importantly, there is no prior evidence as to how restrictions 'react' to changes in the monetary environment, considering that existing studies on monetary policy effects (see, \cite{Piazzesi06}, \cite{Ang11}, and \cite{Orphanides12}) suggest that restrictions selected based on the in-sample process may not be economically plausible around periods of monetary policy shifts, interventions, or under fragile economic conditions. With this in mind, in this paper we study the out-of-sample performance of yields-only DTSMs, in light of the alternative restrictions imposed on the dynamics of risk compensation and attempt to explore whether a real-time Bayesian investor can actually exploit statistical predictability, when making investment decisions. Are DTSMs, which utilise yield curve information only, capable of consistently predicting bond risk premia\footnote{Failure of the EH implies that bond returns are strongly predictable (see, \cite{Fama87}, \cite{Campbell91} and \cite{Cochrane05}, for studies that utilise information coming solely from the yield curve.). In particular, \cite{Fama87} and \cite{Campbell91} propose forward and yield spreads as predictors and suggest that spreads have predictive power on excess returns, while \cite{Cochrane05} use a linear combination of five forward rates as predictors. 
Such evidence, however, is purely statistical.} and generating systematic economic gains to bond investors, out-of-sample? 

Some recent literature (e.g. \cite{Duffee11}, \cite{Barillas11}, \cite{Adrian13}, \cite{Joslin14}), suggests that yields-only DTSMs cannot capture the predictability of bond risk premia, since the required information to predict premia is not spanned\footnote{The spanning hypothesis suggests that the yield curve contains all relevant information required to forecast future yields and excess returns. Unspanned factors are not explained by the yield curve, while at the same time they are useful for predicting risk premia (see, \cite{Cochrane05}, \cite{Duffee11}, \cite{Joslin14} and \cite{Cieslak18}).} by the cross section of yields, implying that more (mainly unspanned) factors are needed. In that respect, \cite{Duffee11} implements a five-factor yields-only DTSM aiming to capture hidden information in the bonds market, while \cite{Wright11}, \cite{Barillas11}, \cite{Joslin14} and \cite{Cieslak15} use measures of macroeconomic activity to predict bond excess returns\footnote{In a recent study however, \cite{BauerHamilton18} cast doubt on prior conclusions, suggesting that the evidence on variables other than the three yield factors predicting excess returns is not convincing.}. In contrast, \cite{Sarno16} and \cite{Feunou18}, implement extended versions of yields-only DTSM and argue that their approaches help those models capture the required predictability, thus, overturning prior evidence. A similar conclusion is reached by \cite{Bauer18}, who studies DTSMs under alternative risk price restrictions. In this study, we attempt to revisit evidently conflicting results on bond excess return predictability based on DTSMs.



Importantly, the above-mentioned studies, either do not consider the out-of-sample economic performance (as in \cite{Duffee11}, \cite{Bauer18}, \cite{Feunou18} and \cite{Giacoletti21}, etc.) or do not fully explore potential economic benefits for bond investors when compared to the non-predictability (constant risk premia) Expectations Hypothesis (EH) benchmark, which is the second empirical question we target in this paper. 
In fact, existing literature on economic value finds evidence of statistical predictability, which nevertheless does not translate into positive economic gains to bond investors (see, \cite{Dellacorte08}, \cite{Sarno16} and recently \cite{Andreasen21}). In particular, using a dynamic mean-variance allocation strategy, framed within a DTSM, \cite{Dellacorte08} and \cite{Sarno16} find that statistical predictability is not turned into superior portfolio performance when compared to the EH benchmark. Consistent results are also presented in \cite{Andreasen21}, in the context of a regime-switching macro-finance term structure model, who suggest that it is difficult to translate evidence of time-variation in expected excess returns into economic benefits to investors.
Qualitatively similar results are found in the literature on economic value generated from  predictive regression models\footnote{Such models are not directly comparable to DTSMs, as the latter make different and stronger modelling assumptions, such as the absence of arbitrage, aiming to explicitly model several aspects of the market (e.g. accurately inferring short-rate expectations and term premia) and obtain further insights on the term structure of risk premia.} on bond excess returns (see for example, \cite{Thornton12}, \cite{Gargano19}, \cite{Bianchi20} and \cite{Wan22}). As in the DTSMs case, the evidence from such studies is conflicting, in some cases pointing towards a negative answer (as in \cite{Thornton12}, \cite{Ghysels18}\footnote{According to \cite{Ghysels18} (and \cite{Wan22}), economic benefits vanish when fully revised macroeconomic information is replaced by real-time data. Our approach is not reliant on macroeconomic data and, as such, our analysis is independent on the debate between 'fully-revised' vs. 'real-time' macros.} and \cite{Wan22}), while in more recent studies  (such as \cite{Gargano19}\footnote{In fact, \cite{Gargano19} find some evidence of economic value for the Fama–Bliss (FB) predictive model. Concurrently, the model by Cochrane–Piazzesi (CP) fails to offer any positive economic gains to bond investors.} and \cite{Bianchi20}) some economic value is retained even for the yields-only case. Motivated in part by the case of predictive regression models, our aim in this paper is to explore, in the context of yields-only DTSMs, whether it is possible to achieve both statistical predictability and economic value by imposing restrictions in the price of risk specification.


\subsection{Sequential Learning and DTSMs with Sparsity}

From a statistical or machine learning viewpoint, imposing restrictions may be thought of as guarding against overfit. If more parameters than needed are used to extract the signal of the market price of risk, it becomes more likely to capture noise rather than systematic patterns, thus leading to poor predictive performance. With this in mind, we propose a novel methodological framework which successfully handles sequential model searches over the space of all possible restrictions in real time, allowing investors to revise their beliefs when new information arrives, thus informing their asset allocation and maximising their expected utility. Setting up in the context of \cite{Bauer18}, we construct a sequential learning scheme following the principles of \cite{Chopin02} and \cite{DelMoral06}. The modelling approach utilises Bayesian inference and forecasting simultaneously, while allowing for model and parameter uncertainty to be incorporated in a sequential manner. We use the developed setup to predict bond excess returns and explore the out-of-sample statistical and economic importance of restrictions. 

Our approach differs from previous studies, allowing us to overcome a number of important challenges and offers several advantages. First, in a similar style to \cite{Wan22} but tailored to the context of DTSMs, it allows us to update the estimates and predictive density as new data arrive, without the need to rerun everything from scratch. Second, it allows for potentially more powerful prediction techniques, such as Bayesian model averaging, to be implemented. In this paper, we develop a sequential version of the stochastic search variable selection (SSVS) scheme (can also be used for Gibbs Variable Selection) that allows incorporating model and parameter uncertainty in a sequential manner. This is of particular importance taking into account that investors often face model uncertainty, which highlights the need for a framework that is capable of monitoring, identifying and adjusting models in real time. Third, it provides a more robust alternative to the Markov Chain Monte Carlo (MCMC) sampler and model choice algorithms of \cite{Bauer18} and \cite{Gargano19}, as its sequential setup naturally provides inference in a parallel way that can potentially overcome issues such as poor mixing, slow convergence properties, and multi-modalities. While such issues do not seem to arise for a given set of restrictions and under the canonical setup of \cite{Joslin11} as in \cite{Bauer18}, the sequential scheme is useful in more challenging setups, such as the exploration of the restriction sets space. 

Our evaluation framework consists of two stages. First, we evaluate the predictive performance using metrics such as the out-of-sample $R^2$ of \cite{Campbell08} ($R_{os}^2$) or the log score (LS) as in \cite{Geweke10}. 
Second, to investigate the economic significance of the out-of-sample excess return forecasts generated by alternative models, we construct a dynamically rebalanced portfolio as in \cite{Dellacorte08} and \cite{Thornton12}, for an investor with power utility preferences, and compute standard metrics 
(see, \cite{Johannes14} and \cite{Gargano19}, among others) in both univariate and multivariate asset allocation setups. 

Our results lead to a host of interesting conclusions regarding the US market. Initially, we confirm \cite{Sarno16} in that yields-only DTSMs with some or no restrictions on the risk premia show evidence of statistical predictability which nevertheless is not translated into systematic economic gains for bond investors. However, we also complement \cite{Sarno16} in that the situation is reversed when heavy restrictions are placed either by sequential model averaging schemes introduced in this paper or by two specific models identified by our framework. The latter are in line with \cite{Cochrane09} and \cite{Duffee11} in that only level risk is priced, but place even heavier restrictions allowing only one or two free risk premia parameters. Those schemes and models offer improved out-of-sample portfolio performance and economically meaningful gains.

\subsection{Outline}

The remainder of this paper is organised as follows. Section 2 describes the modelling framework. Section 3 presents the sequential learning and forecasting procedure along with the framework for assessing the predictive and economic performance of models. Section 4 discusses the data and the sample period used and presents the best models inferred through the sequential SSVS scheme. Section 5 discusses the results both in terms of predictive performance and economic value. Section 6 provides connections with predictive regression models. Finally, Section 7 concludes the paper by providing some relevant discussion.




\section{Dynamic Term Structure Model, Likelihood, and Restrictions}
\label{Model}

In this section we briefly describe the adopted model and the associated likelihood function in order to set up the notation and formulate our research question explicitly. More details can be found in \cite{Joslin11} where this framework was introduced. The model belongs to the no-arbitrage class of Affine Term Structure Models (ATSMs) (see, \cite{Ang03} and \cite{Cochrane05}), under which the one period risk-free interest rate $r_t$\footnote{Working with monthly data implies that $r_t$ is the 1-month yield.} is assumed to be an affine function of an $N \times 1$ vector of state variables $X_t$, namely
\begin{equation}
\label{short rate}
r_{t}=\delta_{0} + \delta_{1}' X_{t}
\end{equation}
where $\delta_{0}$ is a scalar and $\delta_{1}$ is a $N \times 1$ vector. In Gaussian ATSMs, the physical probability measure $\mathbb{P}$ is assumed to be a first-order Gaussian Vector Autoregressive (VAR) process
\begin{equation}
\label{VAR_P}
X_{t}-X_{t-1}= \mu + \Phi X_{t-1} + \Sigma \varepsilon_{t}
\end{equation}
where $\varepsilon_{t} \sim N(0,I_{N})$, $\Sigma$ is an $N \times N$ lower triangular matrix, $\mu$ is a $N \times 1$ vector and $\Phi$ is a $N \times N$ matrix. Lack of arbitrage implies the existence of a pricing kernel $M_{t+1}$, defined as 
\begin{equation}
\label{pricekernel}
M_{t+1} =  \exp( -r_{t}-\frac{1}{2} \lambda_{t}' \lambda_{t} - \lambda_{t}' \varepsilon_{t+1})
\end{equation}
with $\lambda_{t}$ being the time-varying market price of risk which is assumed to be affine in the state $X_t$\footnote{This is the `essentially-affine' specification introduced in \cite{Duffee02}. Existing studies have proposed alternative specifications for the market price of risk, such as the `completely-affine' model of \cite{Dai00}, the `semi-affine' model of \cite{Duarte04}, and the `extended-affine' model of \cite{Cheridito07}. See \cite{Feldhutter16} for a useful comparison of the models.}
\begin{equation}
\label{price of risk}
\lambda_{t} = \Sigma^{-1}\left(\lambda_{0} + \lambda_{1} X_{t}\right)
\end{equation}
where $\lambda_{0}$ is a $N \times 1$ vector and $\lambda_{1}$ is a $N \times N$ matrix. Assuming that the pricing kernel $M_{t+1}$ prices all bonds in the economy and we let $P_{t}^{n}$ denote the time-t price of an n-period zero-coupon bond, then the price of the bond is computed from $P_{t}^{n+1}=E_{t} \left ( M_{t+1}P_{t+1}^{n} \right )$ and leads to the $\mathbb{Q}$ dynamics 
\begin{equation}
\label{VAR_Q}
X_{t}-X_{t-1}= \mu^{\mathbb{Q}} + \Phi^{\mathbb{Q}} X_{t-1} + \Sigma \varepsilon_{t}^{\mathbb{Q}}
\end{equation}
where $\mu^{\mathbb{Q}} = \mu - \lambda_{0}$, $\Phi^{\mathbb{Q}} = \Phi - \lambda_{1}$ and $\varepsilon_{t}^{\mathbb{Q}} \sim N(0,I_{N})$. Define the observed time-t, n-period yield as
\begin{equation}
\label{oldyield_n}
    y_{t}^n = - \frac{\log P_{t}^{n}}{n}.
\end{equation}
The  $J \times 1$ vector of the yields $\{y_t^n\}_{n=1}^J$, denoted by $y_t$, is also an affine function of the state vector
\begin{equation}
\label{oldyield}
y_{t} = A_{n,X}+B_{n,X}'X_{t},
\end{equation}
where the $J \times 1$ loading vector $A_{n,X}$ and the $J \times N$ loading matrix $B_{n,X}$ are calculated using the above recursions, as $A_{n,X}=-A_{n}/n$ and $B_{n,X} = -B_{n}/n$.

In theory, it is possible to specify the likelihood function based on \eqref{VAR_P} and \eqref{oldyield} but, in practice, estimation and identification of these formulations has been proven to be challenging (see, \cite{Ang03}, \cite{Ang07}, \cite{Chib09}, \cite{Duffee12}, \cite{Hamilton12b}, and \cite{Bauer18}), especially if ATSMs are expressed in terms of an unobserved latent $X_t$. Additional restrictions need to be imposed to ensure identifiability, such as the canonical setup of \cite{Joslin11} that is adopted in this paper. More specifically, $X_t$ is assumed to be linearly related to the observed yields, and as such, perfectly priced by the no-arbitrage restrictions. We rotate $X_t$ to match the first $N$ principal components (PCs) of the observed yields
\begin{equation}
\label{PCs}
\mathcal{P}_{t} = W y_t = WA_{n,X} + WB_{n,X}X_{t},
\end{equation} 
with $W$ being the $N \times J$ matrix that contains the PCs' loadings. Following common practice, we consider the case of $N=3$, noting that the first three extracted PCs are typically sufficient to capture most of the variation in the yield curve and often correspond to its level, slope, and curvature respectively \citep{Litterman91}. Statistical inference can proceed using the observations $Y=\{y_t,\mathcal{P}_{t}: t=0,1,\dots,T\}$. The likelihood factorises into two parts stemming from the $\mathbb{P}$ and $\mathbb{Q}$ respectively. In order to specify the latter, henceforth denoted as $\mathbb{Q}$ likelihood, the affine transformation of \eqref{PCs} is applied to \eqref{VAR_Q} to obtain the dynamics of $\mathcal{P}_{t}$ under $\mathbb{Q}$
\begin{equation}
\label{VAR_P_Q}
\mathcal{P}_{t}-\mathcal{P}_{t-1}= \mu_{\mathcal{P}}^{\mathbb{Q}} + \Phi_{\mathcal{P}}^{\mathbb{Q}}\mathcal{P}_{t-1}  + \Sigma_{\mathcal{P}} \varepsilon_{t}^{\mathbb{Q}}
\end{equation}
and, similarly, the yield equation \eqref{oldyield} can be rewritten as a function of $\mathcal{P}_{t}$\footnote{According to \cite{Duffee11}, outside of knife-edge cases, the matrix $(W B_{n,x})$ is invertible, and as such, $\mathcal{P}_{t}$ contains the same information as $X_t$.}
\begin{equation}\label{newyield}
y_{t} = A_{n,\mathcal{P}}+B_{n,\mathcal{P}} \mathcal{P}_{t},
\end{equation}
where  $\mu_{\mathcal{P}}^{\mathbb{Q}}$, $\Phi_{\mathcal{P}}^{\mathbb{Q}}$, $\Sigma_{\mathcal{P}}$, $A_{n,\mathcal{P}}$ and $B_{n,\mathcal{P}}$ are given in Online Appendix A. Note that in \eqref{oldyield} and \eqref{newyield}, yields are assumed to be observed without any measurement error. Nevertheless, an $N$-dimensional observable state vector cannot perfectly price $J>N$ yields, and as such, we further assume that the $J-N$ bond yields used in the estimation are observed with independent $N(0,\sigma_e^2)$ measurement errors. An equivalent way to formulate this is to write
\begin{equation}
\label{Qmodel}
y_{t}=A_{n,\mathcal{P}}+B_{n,\mathcal{P}} \mathcal{P}_{t} + e_{t,n},\;\;\forall n,
\end{equation}
and to consider the dimension of $e_{t,n}$ as effectively being $J-N$.

In order to specify the $\mathbb{P}$ likelihood we note that $\mathbb{P}$ dynamics of $\mathcal{P}_t$ are of equivalent form to \eqref{VAR_P_Q} with $\mu_{\mathcal{P}}^{\mathbb{P}}=\mu_{\mathcal{P}}^{\mathbb{Q}}+\lambda_{0\mathcal{P}}$ and $\Phi_{\mathcal{P}}^{\mathbb{P}}=\Phi_{\mathcal{P}}^{\mathbb{Q}}+\lambda_{1\mathcal{P}}$, where $\lambda_{0\mathcal{P}}$ is a $N \times 1$ vector and $\lambda_{1\mathcal{P}}$ is a $N \times N$ matrix reflecting the market price of risk in $\mathcal{P}_{t}$ terms. We also follow the identification scheme of \cite{Joslin11} (proposition 1), where the short rate is the sum of the state variables, namely $r_t = i X_t$ with $i$ being a vector of ones, and the parameters $\mu^{\mathbb{Q}}$ and $\Phi^{\mathbb{Q}}$ of the $\mathbb{Q}$-dynamics are given as $\mu^{\mathbb{Q}} = [k_{\infty}^{\mathbb{Q}}, 0, 0]$ and $\Phi^{\mathbb{Q}} = \mbox{diag}(g^{\mathbb{Q}})$, where $g^{\mathbb{Q}}$ denotes a $N \times 1$ vector containing the real and distinct eigenvalues of $\Phi^{\mathbb{Q}}$ \footnote{Alternative specifications for the eigenvalues are considered in \cite{Joslin11}; however, real eigenvalues are found to be empirically adequate.}. The joint likelihood (conditional on $\mathcal{P}_{0}$) can now be written as
\begin{equation}\label{likelihood}
f(Y|\theta)=\left\{\prod_{t=0}^T f^{Q}(y_{t}|\mathcal{P}_{t}, k_{\infty}^{\mathbb{Q}}, g^{\mathbb{Q}}, \Sigma, \sigma_{e}^{2})\right\} \times\left\{\prod_{t=1}^T f^{P}(\mathcal{P}_{t}|\mathcal{P}_{t-1}, k_{\infty}^{\mathbb{Q}}, g^{\mathbb{Q}},\lambda_{0\mathcal{P}},\lambda_{1\mathcal{P}}, \Sigma)\right\},
\end{equation}
where the $\mathbb{Q}$-likelihood components, $f^{Q}(\cdot)$, are given by \eqref{Qmodel} and capture the cross-sectional dynamics of the risk factors and the yields, whereas $\mathbb{P}$-likelihood components, $f^{P}(\cdot)$, capture the time-series dynamics of the observed risk factors. The parameter vector is set to  $\theta=(\sigma_{e}^{2},k_{\infty}^{\mathbb{Q}}, g^{\mathbb{Q}},\lambda_{0\mathcal{P}},\lambda_{1\mathcal{P}}, \Sigma)$.

Note that in the case of all entries in $\lambda_{0\mathcal{P}}$, $\lambda_{1\mathcal{P}}$ being non-zero, also known as the maximally flexible model, the mapping between $\theta$ and $\tilde{\theta}=(\sigma_{e}^{2},k_{\infty}^{\mathbb{Q}}, g^{\mathbb{Q}},\mu_{\mathcal{P}}^{\mathbb{P}},\Phi_{\mathcal{P}}^{\mathbb{P}}, \Sigma)$ is 1-1. This allows for the following equivalent likelihood specification
\begin{equation}
\label{likeihood2}
f(Y|\tilde{\theta})=\left\{\prod_{t=0}^T f^{Q}(y_{t}|\mathcal{P}_{t}, k_{\infty}^{\mathbb{Q}}, g^{\mathbb{Q}}, \Sigma, \sigma_{e}^{2})\right\} \times\left\{\prod_{t=1}^T f^{P}(\mathcal{P}_{t}|\mathcal{P}_{t-1}, \mu_{\mathcal{P}}^{\mathbb{P}},\Phi_{\mathcal{P}}^{\mathbb{P}}, \Sigma)\right\}.
\end{equation}
Hence, loosely speaking, under the maximally flexible model the parameters $k_{\infty}^{\mathbb{Q}}, g^{\mathbb{Q}}$ are estimated mainly from the $\mathbb{Q}$ likelihood, in other words based solely on cross-sectional information and without reference to the real-world dynamics\footnote{According to \cite{Joslin11}, the ordinary least squares estimates of parameters $\mu$ and $\Phi$ are almost identical to those estimated using maximum likelihood.}. But if one or more entries of the $\lambda_{0\mathcal{P}}$, $\lambda_{1\mathcal{P}}$ are set to zero, in other words if restrictions are imposed, then the mapping from $\theta$ to $\tilde{\theta}$ is no longer 1-1, allowing only the likelihood specification of \eqref{likelihood} that directly links $\mathbb{Q}$ parameters with time series information. 

Nevertheless, this raises the issue of how to choose between the $2^{N+N^2}$ possible sets of restrictions in the $\lambda_{0\mathcal{P}}$, $ \lambda_{1\mathcal{P}}$ matrices; e.g. in the case of $N=3$ there are $4,096$ distinct sets of restrictions. \cite{Bauer18} suggests using Bayesian model choice, aiming to maximise the model evidence of each restriction specification. In this paper we propose choosing the restriction set with the optimal predictive performance among all possible restriction sets. Models that are optimal in the Bayesian sense, i.e. achieving the highest model evidence, are typically parsimonious and therefore are expected to exhibit good predictive performance. In a related argument, \cite{Fong20} show that model evidence is formally equivalent with exhaustive leave-$p$-out cross-validation combined with the log posterior predictive scoring rule. Hence, it will not be surprising if the same set of restrictions was obtained from both approaches; in fact this is the case in data from the US market as we illustrate in Section \ref{sec:application}. Nevertheless, this is not always guaranteed to be the case and, in situations where different answers are obtained, the predictive performance criterion may be more relevant in the context of DTSMs from an investor's point of view. 


\section{Sequential Estimation, Model Choice, and Forecasting}

In this section we develop a sequential Monte Carlo (SMC) framework for Gaussian ATSMs. We draw from the work of \cite{Chopin02,Chopin04} (see also \cite{DelMoral06}), and make the necessary adaptations to tailor the methodology to the data and models considered in this paper. Furthermore, we extend the framework to allow for sequential Bayesian model choice by incorporating the SSVS algorithm that allows searching over $2^{N+N^2}$ models; see \cite{Schafer13} for some relevant work in the linear regression context. Overall, the developed framework allows the efficient performance of tasks such as sequential parameter estimation, model choice, and forecasting. We begin by providing the main skeleton of the scheme and then provide the details of its specific parts, such as the MCMC scheme for exploring the model space, and the framework for obtaining and evaluating the economic benefits of predictions.

\subsection{Sequential Framework}

Let $Y_{0:t}=(Y_0,Y_1\dots,Y_t)$ denote all the data available up to time $t$, such that $Y_{0:T}=Y$. Similarly, the likelihood based on data up to time $t$ is $f(Y_{0:t}|\theta)$ and is defined in \eqref{likelihood}. Combined with a prior on the parameters $\pi(\theta)$, see Online Appendix B.1 for details, it yields the corresponding posterior 
\begin{equation}\label{posterior}
 \pi(\theta|Y_{0:t})= \frac{1}{m(Y_{0:t})}f(Y_{0:t}|\theta)\pi(\theta),
\end{equation} 
where $m(Y_{0:t})$ is the model evidence based on data up to time t. Moreover, the posterior predictive distribution, which is the main tool for Bayesian forecasting, is defined as
\begin{equation}\label{predictive}
f(Y_{t+h}|Y_{0:t})=\int f(Y_{t+h}|Y_{t},\theta)\pi(\theta|Y_{0:t})d\theta
\end{equation}
where $h$ is the prediction horizon. Note that the predictive distribution in \eqref{predictive} incorporates parameter uncertainty by integrating $\theta$ out according to the posterior in \eqref{posterior}. Usually, prediction is carried out by expectations with respect to \eqref{predictive}, e.g. $E(Y_{t+h}|Y_{0:t})$ but, since \eqref{predictive} is typically not available in closed form, Monte Carlo can be used in the presence of samples from $\pi(\theta|Y_{0:t})$. This process may accommodate various forecasting tasks; for example forecasting several points, functions thereof, and potentially further ahead in the future. A typical forecasting evaluation exercise requires taking all the consecutive times $t$ from the nearest integer of, say, $T/2$ to $T-1$. In each of these times, $Y_{0:t}$ serves as the training sample, and points of $Y$ after $t$ are used to evaluate the forecasts. Hence, carrying out such a task requires samples from \eqref{predictive}, and therefore from $\pi(\theta|Y_{0:t})$, for several times $t$. Note that this procedure can be quite laborious and in some cases infeasible.

An alternative approach that can handle both model choice and forecasting assessment tasks is to use sequential Monte Carlo (see, \cite{Chopin02} and \cite{DelMoral06}) to sample from the sequence of distributions $\pi(\theta|Y_{0:t})$ for $t=0,1,\dots,T$. A general description of the  Iterated Batch Importance Sampling (IBIS) of \cite{Chopin02}'s algorithm, see also \cite{DelMoral06} for a more general framework, is provided in Table \ref{tab:ibis}.
\begin{table}[!ht]
\begin{flushleft}
\hrule
\medskip
{\itshape \small
\vspace{0.2cm}
Initialise $N_{\theta}$ particles by drawing independently $\theta_{i}\sim \pi(\theta)$ with importance weights $\omega_{i}=1$, $i=1,\dots,N_{\theta}$. For $t,\dots,T$ and each time for all $i$:\vspace{0.2cm}
\begin{itemize}
\item[(a)] Calculate the incremental weights 
$$
 u_t(\theta_{i}) =f(Y_{t}|Y_{0:t-1},\theta_{i}) =f\big(Y_{t}|Y_{t-1},\theta_{i})
$$
\item[(b)] Update the importance weights $\omega_{i}$ to $\omega_{i}u_t(\theta_{i})$.
\item[(c)] If some degeneracy criterion (e.g. ESS($\omega$)) is triggered, perform the following two sub-steps:
\begin{itemize}
\item[ (i)] Resampling: Sample with replacement $N_{\theta}$ times from the set of $\theta_{i}$s according to their weights $\omega_i$. The weights are then reset to one.
\item[(ii)] Jittering: Replace $\theta_i$s with $\tilde{\theta}_i$s by running MCMC chains with each $\theta_i$ as input and $\tilde{\theta}_i$ as output.
\end{itemize} 
\end{itemize}}
\medskip
\hrule
\end{flushleft}
\caption{IBIS algorithm}
\label{tab:ibis}
\end{table}
The degeneracy criterion is usually defined through the Effective Sample Size (ESS) which is equal to
\begin{equation} \label{ESS}
ESS(\omega)=\frac{(\sum_{i=1}^{N_{\theta}} \omega_{i})^2}{\sum_{i=1}^{N_{\theta}} \omega_{i}^2}
\end{equation}
and is of the form $ESS(\omega)<\alpha N_{\theta}$ for some $\alpha \in (0,1)$, where $\omega$ is the vector containing the weights. 

The IBIS algorithm provides a set of weighted $\theta$ samples, or else particles, that can be used to compute expectations with respect to the posterior, $E[g(\theta)|Y_{0:t}]$, for all $t$ using the estimator $\sum_{i}[\omega_i g(\theta_{i})]/\sum_{i}\omega_i$. \cite{Chopin04} shows consistency and asymptotic normality of this estimator as $N_{\theta}\rightarrow \infty$ for all appropriately integrable $g(\cdot)$. The same holds for expectations with respect to the posterior predictive distributions, $f(Y_{t+h}|Y_t)$; the weighted $\theta$ samples can be transformed into weighted samples from $f(Y_{t+h}|Y_t)$ by simply applying $f(Y_{t+h}|Y_t,\theta)$. A very useful by-product of the IBIS algorithm is the ability to compute $m(Y_{0:t})=f(Y_{0:t})$, which is the criterion for conducting formal Bayesian model choice. Computing the following quantity in step (a) in Table \ref{tab:ibis} yields a consistent and asymptotically normal estimator of $f(Y_t|Y_{0:t-1})$
\begin{equation}
 M_t = \frac{1}{\sum_{i=1}^{N_{\theta}}\omega_{i}}\sum_{i=1}^{N_{\theta}}\omega_{i}u_{t}(\theta_{i})
\end{equation}
An additional benefit provided by sequential Monte Carlo is that it provides an alternative choice when MCMC algorithms have poor mixing and convergence properties and, in general, is more robust when the target posterior is challenging, e.g. multimodal. Finally, as we demonstrate in Section \ref{sec:application}, the sequential nature of the algorithm allows it to produce informative descriptive output to monitor the evolution of key parameters in time.

In order to apply the IBIS output to models and data in this paper, the following adaptations or extensions are needed. First, the choice of defining the incremental weights in step (a) in Table \ref{tab:ibis}, also known as data tempering, is suitable for getting access to sequences of predictive distributions, needed to assess forecasting performance, but at the same time it is quite prone to numerical stability issues and very low effective sample sizes, in particular early on, that is at the initial time points. This is because the learning rate is typically higher at the beginning, especially when transitioning from a vague prior. An alternative approach that guarantees a pre-specified minimum effective sample size level, and therefore some control over the Monte Carlo error, is to use adaptive tempering; see, for example, \cite{jasra2011}. In order to combine the benefits of both approaches we use a hybrid adaptive tempering scheme which we present in Online Appendix B.2. The idea of this scheme is to use adaptive tempering within each transition between the posteriors based on $Y_{0:t}$ and $Y_{0:t+1}$ for each $t$. Similar ideas have been applied in \cite{Schafer13} and \cite{Kantas2014}.  
Second, and quite crucially in this paper, we extend the framework presented in Section \ref{SSVS} to handle sequential model searches over the space of all possible risk price restrictions. Third, we note that the MCMC sampler, used in sub-step (ii) of step (c) in Table \ref{tab:ibis}, needs to be automated as it will have to be rerun for each time point and particle without the luxury of having initial trial runs, as it is often the case when running a simple MCMC on all the data. The problem is intensified by the fact that the MCMC algorithm used here, developed in \cite{Bauer18}, consists of independence samplers that are known to be unstable. To address this, we utilise the IBIS output and estimate posterior moments to obtain independence sampler proposals; see Online Appendix B.2 for details. Finally, we connect the IBIS output with the construction of a model-driven dynamically rebalanced portfolio of bond excess returns and calculate its economic value.

\subsection{Sequential Model Choice Across Risk Price Restrictions}
\label{SSVS}

As mentioned in Section \ref{Model}, the specification of the market price of risk is conducted via $\lambda_{0\mathcal{P}}$ and $\lambda_{1\mathcal{P}}$. For brevity of further exposition, let $\lambda^{\mathcal{P}}=[\lambda_{0\mathcal{P}},\lambda_{1\mathcal{P}}]$ and $\lambda=\lambda_{1\mathcal{P}}$. If all the entries in $\lambda^{\mathcal{P}}$ are free parameters we get the maximally flexible model. Alternative models have also been proposed in the existing studies, e.g. \cite{Cochrane09} and \cite{Bauer18}, where some of these entries are set to zero. More specifically, in most models the set of unrestricted parameters is usually a subset of $\lambda^{\mathcal{P}}$. A standard approach to facilitating Bayesian model choice is via assigning spike-and-slab priors \citep{Mitchell1988,George1993,Madigan1994} on each of the $\lambda^{\mathcal{P}}_{ij}$s, via the following mixture
\begin{equation}\label{spikeslab}
\lambda^{\mathcal{P}}_{ij}\sim (1-\gamma_{ij})N(0,\tau_{ij}^{(0)})+\gamma_{ij} N(0,\tau_{ij}^{(1)})  
\end{equation}
where $\gamma_{ij}$s are Bernoulli random variables taking zero value if the corresponding $\lambda^{\mathcal{P}}_{ij}$ is small (almost equal to zero) or non-zero value if it is large (significantly different from zero). Hence $\tau_{ij}^{(0)}$ is typically given a very small value, thus forcing the underlying parameter towards zero, while $\tau_{ij}^{(1)}$ is set to a larger value so that the data determine the value of the parameter in question. More specifically for $\tau_{ij}^{(1)}$, we use the Zellner's g-prior as in \cite{Bauer18}, a rather standard choice to prevent over-penalisation of complex models, an issue often referred to as the Lindley's paradox; see Online Appendix B.1 for more details. The $\gamma_{ij}$s are also estimated using MCMC; see Online Appendix B.2 for details. The proportion of the MCMC draws in which each $\gamma_{ij}$ is equal to one provides the posterior probability of the corresponding $\lambda^{\mathcal{P}}_{ij}$ being non-zero, also known as posterior inclusion probability.

We consider two approaches to Bayesian model choice in order to explore its links with predictive performance. The first approach is to implement the spike-and-slab approach on some data used for training purposes in order to select the top models. The sequential algorithm in Table \ref{tab:ibis} is then applied to each of them, without using spike-and-slab priors and with some $\lambda^{\mathcal{P}}_{ij}$ being exactly equal to zero, extracting their predictive distributions and contrasting them with the observed data. Under the second approach, sequential inference on both the models and the parameters is drawn. This is implemented by running a single instance of the sequential algorithm in Table \ref{tab:ibis}, modified to incorporate the SSVS algorithm based on the spike-and-slab priors. In this case, the parameter vector includes the $\gamma_{ij}$s allowing us to calculate the inclusion probabilities, using the particle weights, at each time $t$ based on all the data up to and including $t$. 

This approach offers several advantages in exploring the landscape of the risk price restriction space as we can monitor potential changes in the importance of different $\lambda^{\mathcal{P}}_{ij}$s over time. Moreover, the global search nature of sequential Monte Carlo may be helpful in exploring this landscape across different models. Each $\theta$ particle contains a set of $\gamma_{ij}$s and corresponds to a particular model. The set of $\theta$ particles therefore contains instances of the leading models among the $2^{N+N^2}$ possible ones. Every time resampling and jittering take place, the list of models can be potentially updated giving more focus to the cases with higher weights, or else posterior probability, and potentially depleting the ones with lower weights. Hence it is now less likely to get trapped in local modes when exploring the model space. Finally, this scheme allows combining different models and incorporating model uncertainty into forecasting via model averaging in a sequential manner.  

\newpage
\subsection{Assessing Predictive Performance and Economic Value}
\label{sec:predEconomicValue}

Failure of the EH implies that bond returns are strongly predictable (see, \cite{Fama87}, \cite{Campbell91}, \cite{Cochrane05}, and \cite{Ludvigson09}, among others). In this section, we attempt to revisit conflicting results reported in the existing studies 
(e.g. \cite{Duffee11}, \cite{Barillas11}, \cite{Adrian13}, \cite{Joslin14}, \cite{Sarno16} and \cite{Feunou18}) on the ability of yields-only DTSMs to capture the predictability of risk premia in the US Treasury market. This is done while exploring sequentially the space of restrictions imposed on the dynamics of risk compensation. Furthermore, we attempt to explore whether statistical predictability, if any, can be turned into economic benefits for bond investors.

\subsubsection{Bond Excess Returns:}
\label{sec:R2OS}

The observed continuously compounded excess return of an $n$-year bond is defined as the difference between the holding period return of the $n$-year bond, expressed above in terms of log prices, and the $h$-period yield as
\begin{equation}\label{excess}
rx_{t,t+h}^{n} = -(n-h) y_{t+h}^{n-h} + ny_{t}^{n} - hy_{t}^{h},
\end{equation}
where $y_t^n$ is defined in \eqref{oldyield_n}.
If, instead of taking the observed one, we take
the model-implied yield $y_{t}^{n}$, from equation (\ref{newyield}), we arrive at the predicted excess returns $\widetilde{rx}_{t,t+h}^{n}$ 
%
\begin{equation}\label{fitted} 
\widetilde{rx}_{t,t+h}^{n} 
= A_{n-h} - A_{n} + A_{h} + B'_{n-h} \widetilde{\mathcal{P}}_{t+h} - (B_{n}-B_{h})'\mathcal{P}_{t},
\end{equation}
where $\mathcal{P}_{t}$ is observed and $\widetilde{\mathcal{P}}_{t+h}$ is a prediction from the model. Our developed framework, see Table \ref{tab:ibis}, allows drawing from the predictive distribution of $(\widetilde{\mathcal{P}}_{t+h},\widetilde{rx}_{t,t+h}^{n})$ based on all information available up to time $t$. More specifically, for each $\theta_i$ particle equation \eqref{VAR_P_Q} can be used to obtain a particle of $\widetilde{\mathcal{P}}_{t+h}$, which can then be transformed into a particle of $\widetilde{rx}_{t,t+h}^{n}$ via equation \eqref{fitted}.

The predictive accuracy of bond excess return forecasts is measured in relation to an empirical benchmark. We follow related literature and adopt the EH as this benchmark, which essentially uses historical averages as the optimal forecasts of bond excess returns. This empirical average is
\begin{equation}\label{eh}
\overline{rx}_{t+h}^{n} = \frac{1}{t-h} \sum_{j=1}^{t-h} rx_{j,j+h}^{n}
\end{equation}
We consider two metrics to assess the predictive ability of models considered. First, following \cite{Campbell08}, we compute the out-of-sample $R^{2}$ ($R_{os}^{2}$) as
\begin{equation}
R_{os}^{2} = 1 - \frac{\sum_{s=t_{0}}^{t} ( rx_{s,s+h}^{n} -\widehat{rx}_{s,s+h}^{n})^{2}}{\sum_{s=t_{0}}^{t} (rx_{s,s+h}^{n} -\overline{rx}_{s+h}^{n})^{2}}
\end{equation}
for $\widehat{rx}_{s,s+h}^{n}$ being the mean of the predictive distribution. Positive values of this statistic mean that model-implied forecasts outperform the empirical averages and suggest evidence of time-varying return predictability.
Second, in order to assess the entire predictive distribution offered by our scheme, rather than just point predictions, we use the log score\footnote{The computational details, results and discussion of the log predictive score are presented in Online Appendix D.}; a standard choice among scoring rules with several desirable properties, such as being strictly proper, see for example \cite{Dawid14}. 
These metrics are aggregated over all prediction times ($t_0$ to $t$) and maturities. In order to get a feeling for how large the differences from the EH benchmark are, we report the p-values from the Diebold-Mariano test (see, \cite{Gargano19}) noting that these are viewed as indices rather than formal hypothesis tests. They are based on t-statistics computed taking into account potential serial correlations in the standard errors (see, \cite{NeweyWest1987}).  

\newpage
\subsubsection{Economic Performance of Excess Return Forecasts:}
\label{sec:EconomicValue}

From a bond investor's point of view it is of paramount importance to establish whether the predictive ability of a model can generate economically significant portfolio benefits, out-of-sample. The portfolio performance may also serve as a metric to compare models that impose different sets of restrictions on the price of risk specification. In that respect, our approach is different from \cite{Thornton12} and \cite{Sarno16}, who test the economic significance for an investor with mean-variance preferences\footnote{In fact, \cite{Sarno16}, also use an approximation of the power utility solution. Furthermore, they allow for the variance to be constant (or rolling window) and in-sample, in line with \cite{Thornton12}.} and conclude that statistical significance is not turned into better economic performance, when compared to the EH benchmark. It is more in line with \cite{Gargano19} and \cite{Bianchi20}, who arrive at similar conclusions for models which utilise information coming solely from the yield curve (e.g. yields, forwards, etc.). 
Computationally, it is quite similar to the approach of \cite{Wan22}, tailored to the context of our paper.

We consider a Bayesian investor with power utility preferences,
\begin{equation}\label{utility}
U(W_{t+h}) = U(w_{t}^{n}, rx_{t+h}^{n}) = \frac{W_{t+h}^{1-\gamma}}{1-\gamma}
\end{equation}
where $W_{t+h}$ is an $h$-period portfolio value and $\gamma$ is the coefficient of relative risk aversion. If we let $w_{t}^{n}$ be a portfolio weight on the risky $n$-period bond and $(1-w_{t}^{n})$ be a portfolio weight of the riskless $h$-period bond, then the portfolio value $h$ periods ahead is given as
\begin{equation}\label{portfolio}
W_{t+h} = (1-w_{t}^{n}) \exp (r_{t}^{f}) + w_{t}^{n} \exp(r_{t}^{f} + rx_{t,t+h}^{n})
\end{equation}
where $r_{t}^{f}$ is the risk-free rate, here synonymous with the $h$-period yield. Such an investor maximises her expected utility over h-periods in the future, based on $x_{1:t} = \{rx_{1:t}^{n}, \mathcal{P}_{1:t}\}$
\begin{equation}\label{expectedutility}
E_{t} [U(W_{t+h})|x_{1:t}] = \int U(W_{t+h}) f(W_{t+h}|x_{1:t})dW_{t+h} = \int U(w_{t}^{n}, rx_{t+h}^{n})f(rx_{t+h}^{n}|x_{1:t})drx_{t+h}^{n},
\end{equation}
where $f(rx_{t+h}^{n}|x_{1:t})$ is the predictive density described earlier. 
At every time $t$, our Bayesian learner solves an asset allocation problem getting optimal portfolio weights by numerically solving 
\begin{equation}
\widehat{w}_{t}^{n} = \arg \max \frac{1}{\sum_{i=1}^{N_\theta} \omega_{i}} \sum_{j=1}^{N_{\theta}} \omega_{j}\Bigg\{\frac{[(1-w_{t}^{n})\exp(r_{t}^{f}) + w_{t}^{n}\exp(r_{t}^{f} + \widetilde{rx}_{t,t+h}^{n,j})]^{1- \gamma}}{1- \gamma} \Bigg\},
\end{equation}
with $N_{\theta}$ being the number of particles from the predictive density of excess returns, weighted using importance weights $\omega_i$, $i=1,...,N_{\theta}$, which come from the IBIS algorithm.  

To obtain the economic value generated by each model, we use the resulting optimum weights to compute the CER as in \cite{Johannes14} and \cite{Gargano19}. In particular, for each model, we define the CER as the value that equates the average utility of each model against the average utility of the EH benchmark specification. Denoting realised utility from the predictive model as $\widehat{U}_t = U\left(\widehat{w}_t^n,\{\widetilde{rx}_{t,t+h}^{n,j}\}_{j=1}^{N_{\theta}}\right)$ and realised utility from the EH benchmark as $\overline{U}_t$, we get
\begin{equation}\label{CER}
CER = \left(\frac{\sum_{s=t_0}^t \widehat{U}_s}{\sum_{s=t_0}^t \overline{U}_s}\right)^{\frac{1}{1-\gamma}}-1
\end{equation}
The above can be extended to multi-asset portfolio allocation by taking $w_{t}^{n}$ to be a vector.




\newpage
\section{Data and Models}
\label{sec:application}

The yield data set contains monthly observations of zero-coupon US Treasury yields with maturities of $(1,2,3,4,5,7,10)$-years, spanning the period from January $1985$ to the end of $2018$. In particular, yields used are the unsmoothed Fama-Bliss yields constructed by \cite{Le13}\footnote{We are grateful to Anh Le for generously providing the data set.}.
We consider two samples, one ending at the end of $2007$ and that, as such, precludes the ensuing financial crisis, and a second one which includes the period after the end of $2007$, a period determined by, first, different monetary actions and establishment of unconventional policies and, second, interest rates hitting the zero-lower bound. The first sample has been used by most of the existing studies (see, \cite{Joslin11}, \cite{Joslin14}, \cite{Bauer18}, and \cite{BauerHamilton18}). Following related literature, we choose the starting date avoiding the early $1980$s, a period with evidence of the Fed changing its monetary policy. The post-$2007$ global financial crisis period is excluded from our first sample due to concerns about the capability of Gaussian ATSMs to deal with the zero-lower bound (see, \cite{Kim12} and \cite{Bauer16}). These concerns are explored in the second sample, spanning the period from January $1990$ to end of $2018$, as in \cite{BauerHamilton18}. The post $2007$ period of the second sample coincides with the vast majority of the recent bond predictability literature (see, \cite{Bianchi20}, \cite{Wan22}, \cite{Borup21} and \cite{Li22}), thus allowing direct comparisons. Overall, the data contain different market conditions and monetary policy actions. In the analyses of the two samples the data are split into a warm-up period, where the data are used only for estimation purposes, and a testing period, starting immediately afterwards, where we start evaluating the model predictions while incorporating additional information as the data become available. Specifically, the training periods are $1985-1996$ and $1990-2007$ respectively. The data processing involves extracting the first three principal components from the yield curve, depicted in Figure 1 of Online Appendix C. 

In terms of models, as mentioned in Section \ref{Model}, there are $4,096$ possible distinct sets of risk price restrictions in the case of three factors driving the state variables. The first model considered is the previously mentioned maximally flexible model, ($M_0$), in line with previous empirical studies (e.g. \cite{Duffee11}, \cite{Sarno16}, and \cite{Bauer18}, among others), which places no restrictions on the risk premia parameters. The next three models were the ones suggested by our developed sequential SSVS scheme, as the models with the highest posterior probability, when it was run in the first warm-up period 1985-1996. The outcome of this run, confirmed the argument of sparsity in the market price of risk specification, since the best models only freed one or two risk price parameters. Specifically, the model with the highest posterior probability is the one that allows a single free parameter, $\lambda_{1,2}$, which we denote as $M_1$. This is the parameter which drives variation in the price of level risk due to changes in the slope of the yield curve. The next model, denoted by $M_2$, has two free risk price parameters, $\lambda_{1,2}$ as $M_1$ and also $\lambda_{1,1}$, whereas under $M_3$ only $\lambda_{1,1}$ is free. This suggests that the variation in the price of level risk is driven by changes in the level factor. Finally, we consider Bayesian model averaging with model weights obtained from our developed sequential SSVS scheme. We explore three formulations: the first one ($M_4$) assumes uniform prior distribution over models, where each element of $\lambda^{\mathcal{P}}$ is independently Bernoulli distributed with success probability $0.5$, as in \cite{Bauer18}. The second one ($M_5$) uses a hierarchical prior, namely Beta$(1,1)$-Binomial; see for example \cite{Wilson10} and \cite{Consonni18}. The list of models is completed with $M_6$, a constrained version of $M_4$ with only $\lambda_{1,1}$ and $\lambda_{1,2}$ allowed to be non-zero.

\section{Empirical Results}

This section presents the main results on the statistical and economic performance of excess return forecasts. In particular, we assess models based on different sets of restrictions and explore the evident puzzling behaviour between statistical predictability and meaningful out-of-sample economic benefits for bond investors. Furthermore, we monitor how the optimum set of restrictions behaves around periods of monetary policy shifts, interventions, and fragile economic conditions.

\subsection{Yield Curve and Risk Price Dynamics}
\label{sec:Lambdas}

\begin{figure}[!htbp]
\begin{center}
\includegraphics[trim = 0mm 10mm 0mm 10mm, height=7.8in,center]{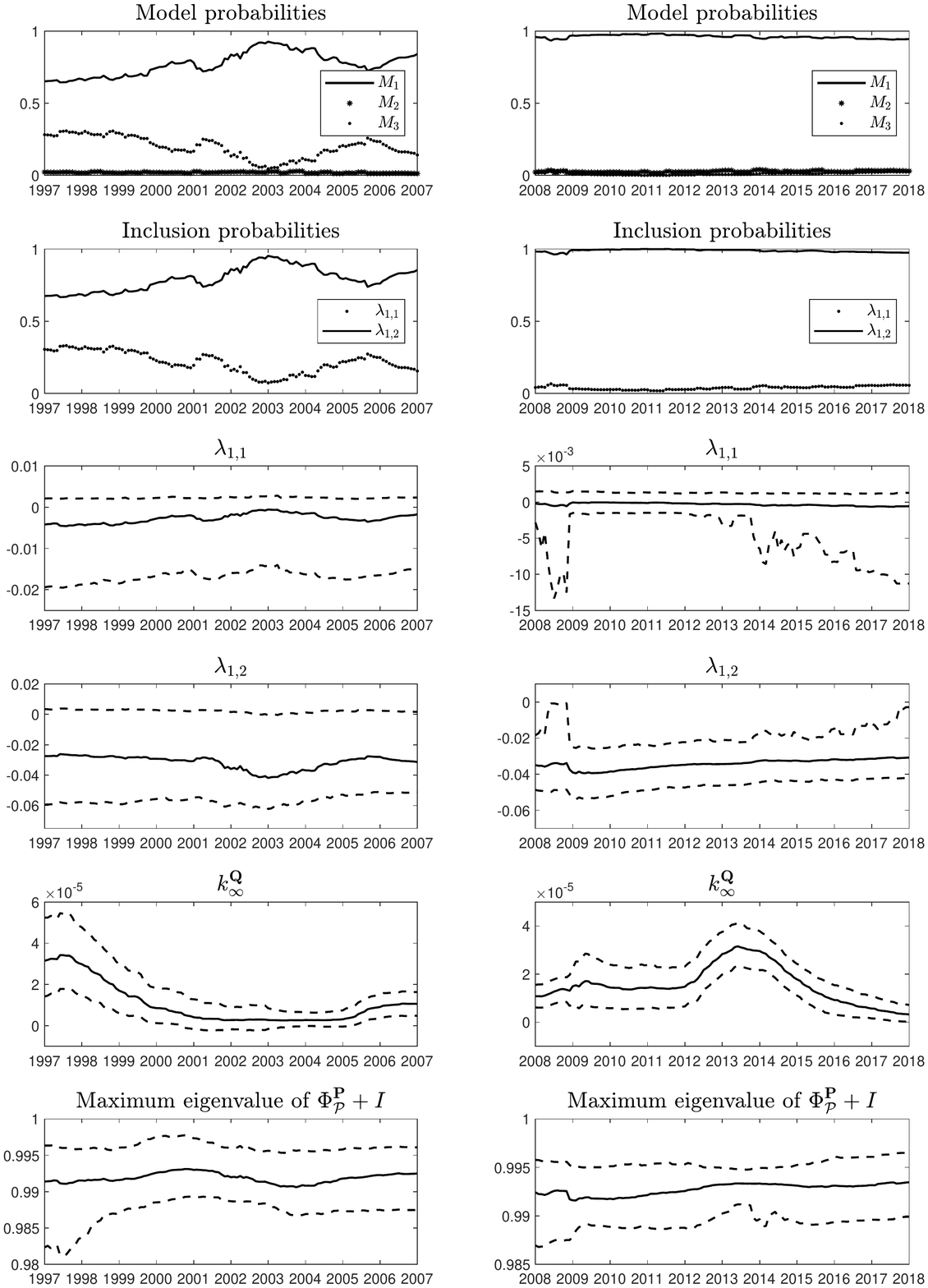}
\caption{\scriptsize Output from model $M_6$ fitted on the first sample (January $1985$ to end of $2007$), left panel, and the second sample (January $1990$ to end of $2018$), right panel. Focus is given on the periods between January $1997$ to end of $2007$  and January $2008$ to end of $2018$ for which predictions of the model were evaluated, but the estimates are based on all the data from January $1985$ and January $1990$, respectively. The first row contains posterior model probabilities for models $M_1$, $M_2$ and $M_3$. The second row contains posterior probabilities of corresponding $\lambda_{1,j}$, $j\in{1,2}$, being non-zero (i.e. inclusion probabilities). The third and fourth rows present posterior means (solid line) and 95\% credible intervals (dashed line) for the market price of risk parameters $\lambda_{1,1}$ and $\lambda_{1,2}$, respectively. Further rows plot real-time estimates of $k_{\infty}^{\mathbb{Q}}$, which is linked to the long-run mean of the short rate, $r_t$, under $\mathbb{Q}$, and the posterior mean of persistence measured via the largest eigenvalue of the feedback matrix.} \label{fig:InclusionProbs19972008}
\end{center}
\end{figure}

The yield curve behaviour through time (see Online Appendix C for more details) is captured from the sequential setup developed in this paper, which allows monitoring variations across time in the estimates of parameters, restrictions, and the importance thereof.
Figure \ref{fig:InclusionProbs19972008} contains the results obtained from fitting the model $M_6$ on the two previously mentioned samples, focusing on the testing periods. The plots depict information for the parameters $\lambda_{1,1}$, $\lambda_{1,2}$, their posterior inclusion probabilities, the associated posterior model probabilities, as well as the highest eigenvalue of $\Phi_{\mathcal{P}}^{\mathbb{P}}$ and $k_{\infty}^{\mathbb{Q}}$ which is linked to the long-run mean of the short rate $r_t$ under $\mathbb{Q}$. From the restrictions parameters, we chose to report only $\lambda_{1,1}$ and $\lambda_{1,2}$ as these were the only parameters with posterior inclusion probabilities not close to zero. In fact, based on the median probability principle that recommends keeping only variables with inclusion probabilities above $0.5$, perhaps only the $\lambda_{1,2}$ parameter should be allowed to be free, thus pointing to model $M_1$. Nevertheless, as we discuss in the remainder of the paper, it may be helpful to consider freeing $\lambda_{1,1}$. 

The posterior inclusion probabilities of the risk premia parameters are rather stable in the second sample, whereas they vary slightly in the first one in line with the changes in policy actions. During the conundrum period, the inclusion probability for $\lambda_{1,1}$ increases while at the same time that for $\lambda_{1,2}$ deteriorates. This suggests that the parameter, which links compensation for level risk to the slope factor, becomes more important during periods of yield curve steepening than periods when the curve flattens. Similarly, the parameter $\lambda_{1,1}$ is more likely to be important during periods where the level of the term structure increases. In terms of model probabilities, $M_1$ performs better than $M_3$, as follows from the posterior inclusions probabilities of $\lambda_{1,2}$ and $\lambda_{1,1,}$ respectively. In the second sample the picture is clearer, as the model selection procedure suggest that only the $\lambda_{1,2}$ parameter should be left free, and consequently $M_1$ is the clear winner.


It is also interesting to look at the long-run mean of the short rate $r_t$ under $\mathbb{Q}$, over time, the posterior trajectory of which follows interest rate expectations and yield curve fluctuations. In the first sample, it starts at a high level and progressively moves down to zero until $2005$. This is followed by an increase during and after the conundrum period of $2004$-$2006$, due to the substantial increase of the federal funds rate. Qualitatively similar conclusions are made when looking at the second sample, where the long-run mean level starts slightly increasing after $2008$, reflecting the steepening of the curve due to the Fed's policies. The increase is more pronounced during and peaks around the $2013$ 'taper tantrum' events, reflecting and capturing the sharp increase in medium-to-long maturity yields, and drops afterwards. Finally, we monitor the largest eigenvalue of the feedback matrix $\Phi_{\mathcal{P}}^{\mathbb{P}}+I$, where $I$ denotes the identity matrix. Its posterior mean remains nearly constant and very close to unity over the entire sample period, indicating a generally high $\mathbb{P}$-persistence, implied by the restrictions imposed on the risk-price parameters. This also reflects an enhanced time variation of short rate expectations and more stable risk premiums, implying a larger role of the expectation component over the risk premium component, in line with policy making and interventions. 

\subsection{Bond Return Predictability}
\label{sec:Predictability}

This subsection presents results on the predictive performance exercise described in Section \ref{sec:R2OS}.
Table $\ref{table:R2OS}$ reports $R_{os}^2$ values for all models across bond maturities, prediction horizons and sub-periods.
Results for the first sample (1985-2007), suggest that the maximally flexible model $M_0$, widely used in the vast majority of prior studies, performs rather poorly out-of-sample compared to the EH benchmark, as showcased by predictive $R_{os}^2$ that are mostly negative, especially at the longer maturities (beyond $5$-year). This is not the case for the second sample (1990-2018) where $M_0$ offers evidence of predictability for investment horizons bigger than $1$ month, generating
scores which are mostly positive and highly significant, especially at the short end of the maturity spectrum and at longer prediction horizons. 

Results are generally better when heavy restrictions are imposed on the risk price dynamics (models $M_1-M_6$),  generating more accurate forecasts, as showcased by positive $R_{os}^2$ values, suggesting strong evidence of out-of-sample bond return predictability. Models $M_1$ and $M_2$ perform quite well in the second sample, which could be partly explained by the parameter $\lambda_{1,2}$ being helpful when the curve steepens. On the other hand model $M_3$, which does not contain $\lambda_{1,2}$, does better in the first sample. This reveals that shocks to the level of the yield curve, captured through $\lambda_{1,1}$, are important components of time-varying risk premia, mainly, during high yield and low uncertainty periods. In terms of the sequential model averaging schemes, they all perform consistently well doing slightly better in the second sample and at long prediction horizons.  

Furthermore, we confirm that predictability is substantially higher following the post-crisis recession period, when the US market experienced high uncertainty and low interest rates, compared to the pre-crisis low volatility and high yield period, where $R_{os}^2$ are substantially lower. This finding is robust across models tested and methodologies applied.

These observations are in contrast to prior literature (see, \cite{Duffee11}, \cite{Barillas11}, \cite{Adrian13}, and \cite{Joslin14}), which suggests that yields-only DTSMs are not capable of capturing the predictability of bond risk premia, and are more in line with the results of \cite{Sarno16} and \cite{Feunou18} who argue that their modelling approaches help models to capture the required predictability of excess returns, in the context of DTSMs. Another message coming out of the analysis, despite the very good performance of models $M_1$-$M_3$, is that sequential model averaging can provide a reliable solution in both time periods. The results based on LS are also qualitatively similar and can be found in Online Appendix D.

\subsection{Economic Performance}
\label{sec:EconomicPerformance}

In this subsection, we concentrate on performance in terms of economic value, as described in Section \ref{sec:predEconomicValue}. We therefore ask whether the predictive ability of models can be exploited by a real-time Bayesian investor when making investment decisions. We initially consider a univariate asset allocation setup which, then is extended to a multivariate allocation framework, where investor jointly models bond excess returns across maturities. To assess the robustness of our results and establish a better link with existing literature, we investigate the economic evidence considering three different scenarios for investors. The first two prevent them from taking extreme positions, while the third relaxes restrictions and allows for maximum leveraging and short-selling. In particular, in the first scenario we follow \cite{Thornton12}, \cite{Sarno16} and \cite{Gargano19} and restrict portfolio weights to range in the interval $[-1, 2]$, thus imposing maximum short-selling and leveraging of 100\% respectively. Second, we follow \cite{Huang20} and restrict portfolio weights to the interval $[-1, 5]$ which keeps maximum short-selling at 100\% while increasing the upper bound, which now amounts to a maximum leveraging of 400\%\footnote{The results and discussion of this allocation scenario are presented in Online Appendix E.1.}. Third, we follow \cite{Bianchi20} and \cite{Wan22} and impose no allocation restrictions to investors, allowing for portfolio weights to be unbounded. Finally, to make our results directly comparable to previous studies, we set the coefficient of risk aversion to $\gamma = 5$, across scenarios.

Table $\ref{table:CER2007}$ reports results for the annualised CER values, generated using out-of-sample forecasts of bond excess returns across maturities and prediction horizons in the first sample. Panel A presents evidence under the first investment scenario. Results show evidence of positive out-of-sample economic benefits, mainly at longer maturities. In particular, we find that, in most models, corresponding CERs are positive and non-negligible, indicating that yields-only models with heavy restrictions on the dynamics of risk compensation, not only provide statistical evidence of out-of-sample predictability, but also generate valuable economic gains for bond investors relative to the EH benchmark. Concurrently, the maximally flexible model $M_0$, fails to offer any positive out-of-sample economic benefits compared to the EH benchmark, generating CER values which are consistently negative across the maturity spectrum and investment horizons. 
Furthermore, consistent with the conclusions coming from the predictability analysis in Section \ref{sec:Predictability}, such models are identified to be the exact same ones, suggesting that only models which allocate one or two non-zero risk price parameters solely to the level factor are able to generate meaningful economic gains for investors who dynamically rebalance their portfolio when new information arrives. Finally, the developed sequential SSVS scheme, which only searches among the best models available, attains very good performance, generating CER values which are quantitatively similar, compared to model $M_1$. Those findings are in contrast to the conclusions of \cite{Thornton12} and \cite{Sarno16}, who argue that bond investors who utilise information from the yield curve only are not able to systematically earn any economic premium out-of-sample.

Importantly, our results suggest larger gains from predictability at longer maturities (beyond 5-years) and investment horizons (beyond 6-months), reflecting substantially higher and statistically significant CER values and consequently more profitable investments. This is in line with prior empirical evidence (see, \cite{Bianchi20}) which states that long-dated yields contain substantial predictive power beyond the first five years. For example, for model $M_1$ at the $6$-month investment horizon, CER is $1.20\%$ ($1.73\%$) for a $7$-year ($10$-year) maturity bond, while at the $12$-month horizon, corresponding CER is $1.13\%$ ($1.70\%$) and highly significant for a $7$-year ($10$-year) maturity bond. Qualitatively similar gains are generated for most of the models. 
In turn, for the sequential SSVS model $M_6$, which searches only among the best models available, at the $6$-month investment horizon, CER is $1.20\%$ ($1.63\%$) for a $7$-year ($10$-year) maturity bond, while at the $12$-month prediction horizon, corresponding CER is $1.08\%$ ($1.50\%$) and highly significant for a $7$-year ($10$-year) maturity bond.
In fact, during the first sub period, those two models appear to outperform all other models tested.

Comparing the performance of alternative model specifications over time sheds light on the importance of particular restrictions across monetary policy actions and market conditions. Table $\ref{table:CER2018}$ displays CER values for the period ending in $2018$, which covers the aftermath of the recession of $2007$-$2009$ as well as the most interesting phases of the unfolding of the Fed's policy responses to it. Results reveal that economic benefits are even more pronounced compared to the pre-crisis period. Such an upturn in CER values occurs across the maturity spectrum with a tendency for substantially larger gains at the long end of the curve, where investments on the $10$-year maturity bond remain the most profitable. 
In particular, looking at all models, other than $M_0$ and $M_3$, CER values for a $10$-year maturity bond almost double when compared to the pre-crisis period. 
For example, for model $M_1$ at the $6$-month investment horizon, CER is up to $2.44\%$ for a $10$-year maturity bond, while at the $12$-month horizon, corresponding CER is $2.22\%$ and highly significant. Qualitatively similar results are observed for other models tested, as displayed in Table $\ref{table:CER2018}$. Importantly, positive and significant gains are now generated at shorter maturities and investment horizons. For example, for model $M_1$ at the $1$-month investment horizon, CER is $1.55\%$ and significant for a $7$-year maturity bond, while for model $M_4$ at the $3$-month horizon, CER is $0.36\%$ and significant for a $4$-year maturity bond.
Finally, models that have been inferred via the sequential SSVS scheme developed (e.g. model $M_4$) do very well offering qualitatively similar CER values to model $M_1$.


Next, we move to investigate asset allocation using the scenario where no allocation restrictions are imposed on investors. Panels B in Tables $\ref{table:CER2007}$ and $\ref{table:CER2018}$ present results for the annualised CER values. Our results in this case, are even more pronounced compared to the constrained allocation scenario of Panel A, since CER values increase substantially, across the maturity spectrum. More specifically, we find that for model $M_1$, at the $3$-month investment horizon, CERs increase up to $3.41\%$ and significant for a $10$-year maturity bond, while for model $M_4$, at the $9$-month horizon, CER value jumps to $3.15\%$.  
Importantly, models with some or no restrictions on the risk price dynamics, such as the maximally flexible model $M_0$, still fail to offer any positive out-of-sample economic benefits compared to the EH benchmark. No matter the bond maturity, the investment horizon and the period considered, generated CERs are consistently negative, revealing no benefits to investors, in line with \cite{Thornton12} and \cite{Sarno16}.

Turning to the multi-asset allocation exercise, results are presented in Table $\ref{table:JOINT}$. Panel A reports annualised CER values for the first sample, while Panel B for the second sample. Our results remain qualitatively similar to the univariate allocation setup. In particular, results reveal that the maximally flexible model $M_0$ continues to fail offering any out-of-sample benefits to investors, generating CERs which are consistently negative across investment horizons and sub-periods. The situation is reversed for models with heavy restrictions on the dynamics of risk compensation (e.g. models $M_1$ and $M_2$) as well as for models inferred via the developed SSVS scheme (e.g. model $M_6$). In particular, results reveal larger gains from predictability at longer investment horizons (beyond $6$-months), where CERs are positive and significant compared to the EH benchmark. In fact, corresponding CERs are higher (in some cases on the order of $1\%$ per annum) compared to the univariate case, suggesting that gains are not limited to specific maturities, as evidenced also in \cite{Bianchi20}. Similar conclusions, yet more pronounced are revealed during the second sample where economic benefits are substantially higher. Interestingly, model $M_3$, which offers positive gains during the first sample, fails to produce any benefits to investors during the second sample, in line with the univariate allocation case. 



\section{Connections with Predictive Regression Models}

Given the improved performance offered by the use of the sequential SSVS approach in DTSMs, it is natural to ask the question whether it can also be of help in similar contexts such as predictive regression models (see, \cite{Fama87}, \cite{Cochrane05}, \cite{Gargano19}, \cite{Bianchi20}, \cite{Wan22}, among many others). We consider inputs from yields-only data, in fact the $\mathcal{P}_t$s (PCs) are viewed as the only inputs, thus leading to the following model 
\begin{equation}\label{rxn}
rx_{t,t+h}^n = a_h + \mathbf{b}_h'\mathcal{P}_{t}+\sigma_h\epsilon_t 
\end{equation}
where $a_h$, $\sigma_h$ are scalars, and $\mathbf{b}_h$ is $N\times1$ vector of the regression coefficients. Connecting with relevant literature, e.g. \cite{Gargano19}, these inputs are closer to the $CP$ factor as they are linear combinations of the yield across maturities; in that paper the $FB$ factors are maturity-specific whereas the $LN$ factor contains macroeconomic information. To maximise relevance with our approach to DTSMs, we consider a variant of the model in \eqref{rxn} paired with a VAR model
\begin{align}
 \label{VAR_b}
    \mathcal{P}_t - \mathcal{P}_{t-1} &= \mu + \Phi\mathcal{P}_{t-1} + \Sigma\varepsilon_t \\
rx_{t,t+h}^n &= a_h + \mathbf{b}_h'\mathcal{P}_{t+h-1}+\sigma_h\epsilon_t \label{eq:PR}
\end{align}
%
where $\mu$, $\Phi$ and $\Sigma$ are defined as before. The above model can provide forecasts for any $h$; in the case of $h=1$ a forecast is obtained directly from \eqref{eq:PR} whereas, for $h>1$, a prediction of $\mathcal{P}_{t+h-1}$ is drawn first from \eqref{VAR_b}. In other words, for $h=1$ we get a standard predictive regression model and, for $h>1$, we incorporate the VAR dynamics present in DTSMs. This allows us to take advantage of potential benefits from sparse VAR formulations while sparsity can also be imposed on the predictive regression coefficients. The exercise is quite challenging from a predictive regression perspective given the yields-only inputs and the time-constant parameters and volatility; e.g. in \cite{Gargano19} such models based on the $CP$ factor fail to generate economic value.

For a given $h$ and information up to time $T$, the model defined by \eqref{VAR_b} and \eqref{eq:PR} can be estimated from the data $\{\mathcal{P}_t\}_{t=0}^T$, $\{rx_{t,t+h}^n\}_{t=0}^{T-h}$. Since the $\{\mathcal{P}_t\}_{t=0}^T$ are assumed to be directly observed, the overall likelihood is given by the product of VAR and predictive regression likelihoods obtained from \eqref{VAR_b} and \eqref{eq:PR} respectively. We proceed by assigning spike and slab priors to all the elements $\mu$, $\Phi$, $a_h$ and $\mathbf{b}_h$, as well as some standard conjugate priors on the remaining parameters, so that a Gibbs sampler is obtained; see Online Appendix F for details. In the presence of that Gibbs sampler, the IBIS algorithm can then be applied as before. It is worth noting that this algorithm is searching over $2^{15}$ models for each $h$ exploring both predictive regression coefficients and VAR parameters.

Table \ref{table:CERSSVSPREDREG} contains figures concerning the economic values generated by the model in \eqref{VAR_b} and \eqref{eq:PR}, using the sequential SSVS approach. In Panel B, corresponding to the second sample, the  model succeeds in generating substantial economic value, mostly for large prediction horizons and small maturities, with results being more pronounced for the case of no portfolio weight restrictions. 
Nevertheless, in Panel A covering the first sample, the results models fail to generate economic value, except for a couple of cases corresponding to unconstrained portfolio weights. Overall, these preliminary results are encouraging and suggest that it would be worthwhile to explore this methodology further, with alternative inputs, time-varying parameters and volatility. 

\newpage
\section{Discussion}

In this paper we focused on the DTSMs, explored their predictability and whether it can translate into economic benefits for investors. Our findings complement \cite{Sarno16} suggesting that economic value can only be obtained, in addition to predictability, if extreme and specific restrictions are placed on their market price of risk specification. In order to implement this approach, we adapted Bayesian variable selection, as in \cite{Bauer18}, to a sequential setting that allows identifying the optimal set of restrictions in real time. The sequential version of the SSVS scheme developed successfully identifies such restrictions either directly, as it can be applied on its own, or indirectly by suggesting specific restrictions that set all risk premia parameters to zero, except for one or two of them ($\lambda_{1,2}$ and potentially $\lambda_{1,1}$). The results are robust to several portfolio allocation scenarios and different time periods, with the performance in the post-$2008$ recession period with no portfolio allocation restrictions being more pronounced, and are driven mostly by long prediction horizons and larger maturities. 

From a statistical viewpoint the problem may be viewed as imposing sparsity. Standard approaches to sparsity include ridge and Lasso regression. However the former is equivalent to assigning normal priors on the risk premia parameters of the maximally flexible model and was implemented without success, whereas the Bayesian versions of the latter, which are essential to conduct portfolio allocation, are generally not associated with sparsity. Instead the use of spike and slab priors is one of the default approaches to impose sparsity in the Bayesian context. It may be useful to explore alternative options; see for example \cite{Polson11} and the references therein. In terms of identified optimal restrictions, our findings are in line with \cite{Cochrane09} and \cite{Duffee11} in that only the level risk is priced, but our adopted models impose further restrictions. The findings of the empirical analysis suggest that these additional restrictions are necessary to produce economic value. An alternative market price of risk specification would be the reduced rank approach of \cite{Joslin11}, perhaps not directly, being less strict than the restrictions implied \cite{Duffee11}, but paired with spike and slab priors. The sequential SSVS approach may also be useful in the context of predictive regressions, aiming to offer improved economic value to existing approaches.


Furthermore, our results reveal some evidence of time variation in the parameters and restrictions; for the latter this is only viewed in the first sample. The sequential SSVS approach can capture such variations to some extent but this is also closely linked to the choice of the data window. Using an expanding window, as in this paper, could work well for small and perhaps moderate changes, but is unlikely to capture extreme shocks such as the Covid-19 period where a shorter window would be more appropriate. Going forward, it would be interesting to consider DTSM models with time varying parameters, for example the mean under the pricing measure appears to be time varying as suggested from the output offered by the IBIS algorithm, or regime switching approaches tailoring for example \cite{Andreasen21} to the context of DTSMs. Another promising future direction is to incorporate spanned or unspanned macroeconomic variables in the models; see, for example, \cite{Joslin14}. 


\bibliographystyle{informs2014}

\bibliography{referencefile3}
\newpage



\newpage

\begin{table}[!htbp]\scriptsize						\caption{Out-of-sample statistical performance of Bond excess return forecasts measured via $R_{os}^2$.
}
\label{table:R2OS}
\begin{center}
\begin{tabular}{llllllllllllll}
\midrule \midrule
\multicolumn{7}{c}{\bf Panel A: Period - 1985 - 2007} &         \multicolumn{7}{c}{\bf Panel B: Period - 1990 - 2018} \\
\midrule
\multicolumn{1}{m{0.5cm}}{\bf h}
&\multicolumn{1}{m{0.8cm}}{\bf 2Y} &\multicolumn{1}{m{0.8cm}}{\bf 3Y} & \multicolumn{1}{m{0.8cm}}{\bf 4Y} & \multicolumn{1}{m{0.5cm}}{\bf 5Y} & \multicolumn{1}{m{0.5cm}}{\bf 7Y} & \multicolumn{ 1}{m{0.75cm}}{\bf 10Y} & \multicolumn{1}{m{0.5cm}}{\bf h} &\multicolumn{1}{m{0.8cm}}{\bf 2Y} &\multicolumn{1}{m{0.8cm}}{\bf 3Y} & \multicolumn{1}{m{0.8cm}}{\bf 4Y} & \multicolumn{1}{m{0.8cm}}{\bf 5Y} & \multicolumn{1}{m{0.8cm}}{\bf 7Y} & \multicolumn{1}{m{0.8cm}}{\bf 10Y}\\
\midrule                                             \multicolumn{14}{c}{$\mathbf{M_0}$}  \\
\midrule
\bf 1m &        0.03*** &     0.01** &     0.00** &     -0.02* &       -0.03 &       -0.08 &         \bf 1m &       -0.02* &     -0.02 &     -0.03 &     -0.02 &       -0.01 &       -0.02* \\

\bf 3m &       0.06** &    0.02** &    0.01** &    -0.04** &     -0.06** &       -0.22 &         \bf 3m &      0.08*** &    0.05** &    0.02** &    0.02** &     -0.01* &       -0.02** \\

\bf 6m &       0.09** &    0.04** &    0.01** &    -0.04** &    -0.14* &      -0.35 &         \bf 6m &      0.26*** &    0.20*** &    0.13*** &    0.10** &    0.06**  &      0.03** \\

\bf 9m &       0.07* &    0.02* &    -0.02* &    -0.07* &    -0.20 &     -0.43 &         \bf 9m &      0.39*** &    0.28***   &    0.19** &   0.16** &    0.14** &     0.09** \\

\bf 12m &         0.08* &    0.02 &    -0.03 &    -0.06 &    -0.22 &     -0.43 &        \bf 12m &         0.48*** &    0.40*** &    0.28** &    0.25** &    0.21** &     0.16** \\
\midrule                                               \multicolumn{14}{c}{$\mathbf{M_1}$} \\
\midrule
        \textbf{1m} &    0.05*** &     0.04*** &     0.03*** &     0.03** &       0.02** &       0.03** &         \bf 1m &       0.03*** &     0.05*** &     0.04*** &     0.04** &       0.03** &       0.04*** \\

        \textbf{3m} &    0.07** &    0.06** &    0.06** &    0.04* &     0.06** &       0.05**  &         \bf 3m &      0.10*** &    0.12*** &    0.10*** &    0.10*** &     0.09*** &       0.12*** \\

  \textbf{6m} &    0.10** &    0.09* &    0.10** &    0.09* &    0.09* &     0.08**  &         \bf 6m &      0.19*** &    0.23*** &    0.20*** &    0.18*** &    0.17*** &     0.20*** \\

 \textbf{9m} &    0.08 &    0.08 &    0.10* &    0.10* &    0.11* &       0.09*   &         \bf 9m &      0.28*** &    0.35*** &    0.34*** &    0.32*** &    0.28*** &       0.32*** \\

\textbf{12m} &    0.07 &    0.06 &    0.08 &    0.08 &     0.09* &     0.08**  &        \bf 12m &         0.20** &    0.36*** &    0.40*** &    0.39*** &     0.37*** &     0.40***  \\
\midrule 
\multicolumn{14}{c}{$\mathbf{M_2}$} \\
\midrule 
        \textbf{1m} &    0.04*** &     0.03** &       0.03** &      0.02* &       0.02* &       0.03** &         \bf 1m &       0.04*** &     0.05*** &       0.04*** &      0.03*** &       0.03** &       0.03*** \\

    \textbf{3m} &    0.07** &     0.06** &      0.06** &       0.04* &       0.05* &       0.05** &         \bf 3m &      0.12*** &     0.12*** &      0.09*** &       0.09*** &       0.08*** &       0.11*** \\

\textbf{6m} &    0.09** &    0.08* &     0.09* &     0.08* &     0.08* &      0.07* &         \bf 6m &      0.24*** &    0.25*** &     0.21*** &     0.18*** &     0.16*** &      0.18*** \\

 \textbf{9m} &    0.07 &    0.07 &     0.09 &     0.08 &     0.08* &     0.06* &         \bf 9m &      0.35*** &    0.39*** &     0.36*** &     0.33*** &     0.28*** &     0.30*** \\

\textbf{12m} &    0.05 &       0.04 &       0.05 &       0.05 &       0.06* &       0.03*  &        \bf 12m &         0.30*** &       0.42*** &       0.43*** &       0.42*** &       0.38*** &       0.38*** \\
\midrule                                               \multicolumn{14}{c}{$\mathbf{M_3}$} \\
\midrule
        \textbf{1m} &    0.03** &    0.02* &      0.02* &     0.02* &       0.02* &       0.02* &         \bf 1m &       0.03*** &    0.00* &      0.00 &     0.00 &       0.00 &       -0.02 \\

 \textbf{3m} &    0.05** &     0.05** &       0.05** &       0.03** &       0.05** &       0.04** &         \bf 3m &      0.10*** &     0.03** &       0.01*  &       0.01  &       0.00   &       -0.05 \\

 \textbf{6m} &    0.09** &    0.07** &     0.07** &       0.06** &       0.06*** &       0.05***  &         \bf 6m &      0.19*** &    0.07*** &     0.03** &       0.01*  &       0.01   &       -0.04 \\

\textbf{9m} &    0.13** &     0.12** &     0.11** &       0.10*** &       0.10*** &       0.07***  &         \bf 9m &      0.29*** &     0.12***  &     0.02**  &       0.00*   &       -0.01 &       -0.06 \\

\textbf{12m} &    0.16** &       0.13** &       0.13** &     0.11*** &       0.11*** &      0.08***  &        \bf 12m &         0.41*** &       0.22***   &       0.07** &     0.03*  &       0.00  &      -0.06 \\
\midrule                                               \multicolumn{14}{c}{$\mathbf{M_4}$} \\
\midrule
        \textbf{1m} &    0.05** &     0.03*** &      0.03** &     0.02** &       0.02** &      0.02** &         \bf 1m &       0.04** &     0.05*** &      0.04*** &     0.03** &       0.03** &      0.04*** \\

 \textbf{3m} &    0.08** &     0.07** &       0.07** &       0.05** &       0.07** &      0.06** &         \bf 3m &      0.13***  &     0.13*** &       0.10*** &       0.10*** &       0.09*** &      0.11*** \\

\textbf{6m} &    0.10** &     0.09** &      0.10** &       0.09** &       0.09** &      0.08*  &         \bf 6m &      0.26*** &     0.27*** &      0.23***  &       0.20*** &       0.18*** &      0.20*** \\

 \textbf{9m} &    0.09 &     0.09* &      0.10* &       0.10*  &       0.11* &       0.11*  &         \bf 9m &      0.37*** &     0.42***  &      0.39***  &       0.35***  &       0.31*** &       0.33*** \\

\textbf{12m} &    0.08 &       0.06 &       0.08 &      0.08*  &      0.10*  &      0.10**  &        \bf 12m &         0.31** &       0.45*** &       0.46*** &      0.45***  &      0.41***  &      0.42*** \\
\midrule                                               \multicolumn{14}{c}{$\mathbf{M_5}$} \\
\midrule
        \textbf{1m} &     0.05*** &       0.04*** &       0.03*** &     0.03** &       0.03** &       0.02** &         \bf 1m &       0.03** &       0.04** &       0.03** &     0.03** &       0.03** &       0.03*** \\

\textbf{3m} &    0.07** &    0.06** &     0.06** &      0.04* &      0.06** &      0.05* &         \bf 3m &      0.11*** &    0.12*** &     0.10*** &      0.10*** &      0.09*** &      0.11*** \\

\textbf{6m} &    0.08* &    0.07* &    0.08* &    0.07* &    0.08* &       0.07*  &         \bf 6m &      0.20*** &    0.24*** &    0.21*** &    0.19*** &    0.17*** &       0.19*** \\

\textbf{9m} &    0.08 &    0.08 &    0.09 &    0.09 &    0.09 &      0.10  &         \bf 9m &      0.30*** &    0.37*** &    0.35*** &    0.33*** &    0.29*** &      0.32*** \\

\textbf{12m} &    0.08 &    0.05 &    0.06 &    0.07 &    0.08 &      0.09  &        \bf 12m &         0.22*** &    0.38*** &    0.41*** &    0.41*** &    0.39*** &      0.41*** \\
\midrule                                               \multicolumn{14}{c}{$\mathbf{M_6}$} \\
\midrule
        \textbf{1m} &    0.04*** &     0.04*** &     0.03*** &     0.03** &       0.02** &       0.03*** &         \bf 1m &       0.04** &     0.05*** &     0.04*** &     0.04*** &       0.04** &       0.04*** \\

 \textbf{3m} &    0.08** &    0.07** &    0.07** &    0.05* &      0.06** &       0.06** &         \bf 3m &      0.10*** &    0.12*** &    0.09*** &    0.10*** &      0.09*** &       0.11*** \\

 \textbf{6m} &    0.10** &    0.10** &    0.10** &    0.09* &    0.09** &      0.09**  &         \bf 6m &      0.20*** &    0.23*** &    0.20*** &    0.18*** &    0.17*** &      0.19*** \\

\textbf{9m} &    0.10* &    0.10* &    0.11* &    0.11* &    0.11* &     0.10*  &         \bf 9m &      0.28*** &    0.34*** &    0.33*** &    0.31*** &    0.28*** &     0.31*** \\

\textbf{12m} &    0.09 &    0.08 &    0.09 &    0.09* &     0.10* &     0.08**  &        \bf 12m &         0.21** &    0.36*** &    0.39*** &    0.39*** &     0.37*** &     0.40***  \\
\midrule \midrule
\end{tabular}
\end{center}
\noindent{\caption*{\scriptsize This table reports $R_{os}^2$ across alternative models, at different prediction horizons, of $h$= 1-month, 3-month, 6-month, 9-month and 12-month. The seven forecasting models used are ATSMs with alternative risk price restrictions. The values are generated using the $R_{os}^2$ measure of \cite{Campbell08}. In particular, $R_{os}^2$ measures the predictive accuracy of bond excess return forecasts relative to the EH benchmark. The EH implies the historical mean being the optimal forecast of excess returns. Positive values of this statistic imply that the forecast outperforms the historical mean forecast and suggests evidence of time-varying return predictability. Statistical significance is measured using a one-sided Diebold-Mariano statistic computed with Newey-West standard errors. * denotes significance at 10\%, ** significance at 5\% and *** significance at 1\% level. The in-sample period for results in Panel A is January 1985 to end of 1996, and the out-of-sample period starts in January 1997 and ends at the end of 2007. The in-sample period for results
in Panel B is January 1990 to end of 2007, and the out-of-sample period starts in January 2008 and ends at the end of 2018.}}
\end{table}


\begin{table}[!htbp]\scriptsize						\caption{Out-of-sample Economic performance of Bond excess return forecasts across prediction horizons - Period: January 1985 - end of 2007.}
\label{table:CER2007}
\begin{center}
\begin{tabular}{llllllllllllll}
\midrule \midrule                                       \multicolumn{7}{c}{\bf Panel A: w= $[-1, 2]$} &         \multicolumn{7}{c}{\bf Panel B: w= $[-\infty,+\infty]$} \\
\midrule
\multicolumn{1}{m{0.5cm}}{\bf h} &\multicolumn{1}{m{0.5cm}}{\bf 2Y} &\multicolumn{1}{m{0.5cm}}{\bf 3Y} & \multicolumn{1}{m{0.5cm}}{\bf 4Y} & \multicolumn{1}{m{0.5cm}}{\bf 5Y} & \multicolumn{1}{m{0.5cm}}{\bf 7Y} & \multicolumn{ 1}{m{1cm}}{\bf 10Y} & \multicolumn{1}{m{0.5cm}}{\bf h} &\multicolumn{1}{m{0.9cm}}{\bf 2Y} &\multicolumn{1}{m{0.9cm}}{\bf 3Y} & \multicolumn{1}{m{0.9cm}}{\bf 4Y} & \multicolumn{1}{m{0.9cm}}{\bf 5Y} & \multicolumn{1}{m{0.9cm}}{\bf 7Y} & \multicolumn{1}{m{0.9cm}}{\bf 10Y}\\
\midrule                                             \multicolumn{14}{c}{$\mathbf{M_0}$}  \\
\midrule
\bf 1m &       0.07 &       -0.39 &        -0.48 &        -1.20 &         -2.01 &          -7.13 &         \bf 1m & -0.31 &       -1.54 &        -1.25 &        -2.11 &         -2.61 &          -7.19 \\

\bf 3m &      -0.40 &      -0.91 &      -1.11 &      -1.51 &        -2.60 &          -6.42 &         \bf 3m &  -0.86 &      -1.61 &      -1.38 &      -1.81 &        -2.86 &          -6.50 \\

\bf 6m &      -0.51 &      -1.15 &      -1.46 &      -1.74 &       -2.74 &        -4.65 &         \bf 6m & -0.87 &      -1.40 &      -1.42 &      -1.65 &       -2.83 &        -4.72 \\

\bf 9m &      -0.59 &      -1.12 &      -1.48 &      -1.72 &       -2.62 &         -3.84 &         \bf 9m & -1.89 &      -2.07 &      -1.89 &      -1.91 &       -2.77 &         -3.85 \\

\bf 12m &         -0.41 &          -0.83 &         -1.17 &         -1.32 &         -2.32 &          -3.32 &         \bf 12m & -2.12 &          -1.98 &         -1.69 &         -1.48 &         -2.44 &          -3.33 \\
\midrule
\multicolumn{14}{c}{$\mathbf{M_1}$}  \\
\midrule
\bf 1m &      0.04 &       0.03 &       0.01 &       0.17 &        1.06 &          0.96 &         \bf 1m & 1.27 &       0.37 &       0.61 &       0.94 &         1.20 &          1.05 \\

\bf 3m &      0.00 &      0.00 &       -0.11 &       0.06 &        0.69 &         0.69 &         \bf 3m & 1.15 &      0.75 &       0.89 &       1.05 &        1.02 &         0.82 \\

\bf 6m &      0.00 &      0.02 &      0.00 &      0.17 &      1.20* &       1.73** &         \bf 6m & 1.41* &      1.49* &      1.65** &      1.75** &      1.91** &       1.87** \\

\bf 9m &      0.00 &      0.00 &      -0.03 &      0.10 &       1.28** &       1.97** &         \bf 9m & 0.77 &      1.31* &      1.69** &     1.83*** &       2.15*** &       2.08*** \\

\bf 12m &         0.00 &         0.00 &        -0.02 &         0.07 &         1.13*** &         1.70*** &         \bf 12m & 0.03 &         0.75 &         1.17* &         1.47** &         1.82*** &         1.81*** \\
\midrule
\multicolumn{14}{c}{$\mathbf{M_2}$} \\
\midrule
\bf 1m &          0.00 &          0.00 &         0.02 &          0.27 &          1.15 &         0.37 &         \bf 1m & 1.06 &          0.01 &          0.17 &          0.58 &          0.70 &          0.66 \\

\bf 3m &          0.00 &          -0.01 &         -0.10 &          0.15 &          0.74 &         0.61 &         \bf 3m & 1.35 &         0.98 &          1.05 &          1.17 &          0.97 &         0.74 \\

\bf 6m &         0.00 &         0.00 &        -0.05 &       0.02 &       1.08** &       1.59** &         \bf 6m & 1.21* &         1.35* &        1.51** &       1.58** &       1.69** &       1.69** \\

\bf 9m &          0.00 &         0.00 &        -0.03 &       -0.08 &       1.11*** &      1.78*** &         \bf 9m & 0.65 &         1.15* &        1.50** &       1.58*** &       1.87*** &      1.84*** \\

\bf 12m &        0.00 &        0.00 &        -0.02 &         -0.11 &     0.90*** &        1.51*** &         \bf 12m & -0.10 &        0.62 &        1.03* &         1.27** &     1.58*** &        1.58*** \\
\midrule
\multicolumn{14}{c}{$\mathbf{M_3}$} \\
\midrule
\bf 1m &      0.00 &         0.03 &        0.26 &         0.60 &          1.84 &          1.02 &         \bf 1m & 1.02 &         1.30 &        1.65 &         2.01 &          1.99 &          1.00 \\

\bf 3m &         0.00 &         0.00 &         0.20* &         0.55* &         0.95 &          0.24 &         \bf 3m & 0.99 &         1.08* &         1.31* &         1.44** &         1.02* &          0.24 \\

\bf 6m &        0.00 &         0.00 &          0.04 &        0.12 &       0.75** &      0.70** &         \bf 6m & 0.74 &         0.91* &          0.93** &        0.90** &       0.84** &      0.70** \\

\bf 9m &        0.00 &        0.00 &         0.01 &      0.02 &      0.67** &      0.68** &         \bf 9m & 0.70 &        0.91* &         0.93** &      0.76** &      0.79** &      0.68** \\

\bf 12m &       0.00 &        0.00  &       0.01* &        0.09** &         0.66** &      0.63 &         \bf 12m & 0.38 &        0.89 &        0.95** &        0.83* &          0.80** &      0.63 \\
\midrule
\multicolumn{14}{c}{$\mathbf{M_4}$} \\
\midrule
\bf 1m &         0.09 &          -0.05 &          -0.44 &          -0.25 &         0.71 &          -0.57 &         \bf 1m & 0.94 &          0.12 &          0.18 &          0.34 &         0.35 &          -0.57 \\

\bf 3m &         0.04 &        -0.02 &         -0.08 &     0.15 &         0.72 &          0.22 &         \bf 3m & 1.26 &        0.91 &         1.07 &    1.24* &         1.02 &          0.22 \\

\bf 6m &        0.03 &         0.04 &         -0.01 &         0.18 &      1.00 &      1.02 &         \bf 6m & 1.03 &         1.00 &          1.13* &         1.20* &      1.24 &      1.02 \\

\bf 9m &        0.00 &         -0.01 &          -0.10 &       0.08 &      1.12* &      1.22** &         \bf 9m & 0.67 &         0.93* &          1.15** &       1.20** &      1.35* &      1.22** \\

\bf 12m &        0.00 &        -0.05 &        -0.14 &         0.00 &        0.95* &      1.01**  &         \bf 12m & 0.17 &        0.58 &        0.79* &         0.96** &        1.09* &      1.01** \\
\midrule
\multicolumn{14}{c}{$\mathbf{M_5}$} \\
\midrule
\bf 1m &         0.05 &       0.20*  &         0.02 &          -0.05 &       0.80 &          0.17 &         \bf 1m & 0.79 &       0.33 &         0.61 &          0.87 &          1.24 &          0.15 \\

\bf 3m &        0.00 &        -0.01 &          -0.06 &     0.11 &          0.57 &          -0.23 &         \bf 3m & 0.65 &        0.33 &          0.45 &     0.61 &          0.52 &          -0.23 \\

\bf 6m &         0.00 &          0.01 &         -0.10 &          -0.03 &         0.65 &        0.58 &         \bf 6m & 0.55 &          0.52 &         0.57 &          0.67 &         0.71 &         0.58 \\

\bf 9m &        0.00 &         0.00 &         -0.10 &          -0.02 &         0.70 &         0.71 &         \bf 9m & 0.46 &         0.57 &         0.61 &          0.62 &         0.74 &         0.71 \\

\bf 12m &        0.00 &        0.00 &          -0.08 &        0.00 &          0.59 &         0.58 &         \bf 12m & 0.35 &        0.53 &          0.50 &        0.60 &          0.62 &         0.58 \\
\midrule
\multicolumn{14}{c}{$\mathbf{M_6}$} \\
\midrule
\bf 1m &      0.00 &       0.00 &       -0.01 &       0.12 &         1.32* &          1.21 &         \bf 1m & 1.41 &       0.76 &       1.00 &       1.33 &         1.42 &          1.22 \\

\bf 3m &      0.00 &      0.00 &       -0.01 &       0.28 &         0.93 &          0.70 &         \bf 3m & 1.51* &      1.08 &       1.16 &       1.28 &         1.11 &          0.78 \\

\bf 6m &      0.00 &      0.00 &      -0.01 &      0.15 &       1.20** &         1.63** &         \bf 6m & 1.45** &      1.44** &      1.54** &      1.59** &       1.69** &         1.66** \\

\bf 9m &      0.00 &      0.00 &       0.00 &    0.08 & 1.29*** &        1.81*** &         \bf 9m & 0.95 &      1.34** &      1.62*** &      1.67*** &       1.91*** &        1.82*** \\

\bf 12m &         0.00 &         0.00 &         0.00 &         0.01 &         1.08*** &         1.50*** &         \bf 12m & 0.19 &         0.80 &         1.12** &         1.32*** &         1.57*** &         1.52*** \\
\midrule \midrule
\end{tabular}
\end{center}
\noindent{\caption*{\scriptsize This table reports annualised certainty equivalent returns (CERs) across alternative models, at different prediction horizons, of $h$= 1-month, 3-month, 6-month, 9-month and 12-month. The coefficient of risk aversion is $\gamma=5$. Panel A presents CERs under the scenario where investors are prevented from extreme investments and as such, portfolio weights are restricted to range in the interval $[-1, 2]$, thus imposing maximum short-selling and leveraging of 100\% respectively. Panel B, reports CER values under the scenario where no allocation restrictions are imposed to investors and, as such, portfolio weights are unbounded, thus allowing for maximum leveraging and short-selling. CERs are generated by out-of-sample forecasts of bond excess returns and are reported in \%. At every time step, $t$, an investor with power utility preferences, evaluates the entire predictive density of bond excess returns and solves the asset allocation problem, thus optimally allocating her wealth between a riskless bond and risky bonds with maturities 2, 3, 4, 5, 7 and 10-years. CER is, then, defined as the value that equates the average utility of each alternative model against the average utility of the EH benchmark. The seven forecasting models used are ATSM with alternative risk price restrictions. Positive values indicate that the models perform better than the EH benchmark. Statistical significance is measured using a one-sided Diebold-Mariano statistic  computed with Newey-West standard errors. * denotes significance at 10\%, ** significance at 5\% and *** significance at 1\% level. The in-sample period is January 1985 to end of 1996, and the out-of-sample period starts in January 1997 and ends at the end of 2007.}}
\end{table}

\begin{table}[!htbp]\scriptsize \caption{Out-of-sample Economic performance of Bond excess return forecasts across prediction horizons - Period: January 1990 - end of 2018.}
\label{table:CER2018}
\begin{center}
\begin{tabular}{llllllllllllll}
\midrule \midrule                                       \multicolumn{7}{c}{\bf Panel A: w= $[-1, 2]$} &         \multicolumn{7}{c}{\bf Panel B: w= $[-\infty,+\infty]$} \\
\midrule
\multicolumn{1}{m{0.5cm}}{\bf h} &\multicolumn{1}{m{0.5cm}}{\bf 2Y} &\multicolumn{1}{m{0.5cm}}{\bf 3Y} & \multicolumn{1}{m{0.5cm}}{\bf 4Y} & \multicolumn{1}{m{0.5cm}}{\bf 5Y} & \multicolumn{1}{m{0.5cm}}{\bf 7Y} & \multicolumn{ 1}{m{0.9cm}}{\bf 10Y} & \multicolumn{1}{m{0.5cm}}{\bf h} &\multicolumn{1}{m{0.9cm}}{\bf 2Y} &\multicolumn{1}{m{0.9cm}}{\bf 3Y} & \multicolumn{1}{m{0.9cm}}{\bf 4Y} & \multicolumn{1}{m{0.9cm}}{\bf 5Y} & \multicolumn{1}{m{0.9cm}}{\bf 7Y} & \multicolumn{1}{m{0.9cm}}{\bf 10Y}\\
\midrule                                             \multicolumn{14}{c}{$\mathbf{M_0}$}  \\
\midrule
\bf 1m &       -0.91 &       -1.41 &        -2.45 &        -2.85 &         -2.23 &          -5.12 &  \bf 1m & -4.27 &       -4.68 &        -4.39 &        -3.79 &         -2.31 &          -5.92 \\

\bf 3m &      -0.21 &      -0.18 &      -0.59 &      -0.69 &        -0.77 &          -2.25 &    \bf 3m & -0.82 &      -1.12 &      -1.01 &      -0.54 &        -0.56 &          -2.03 \\

\bf 6m &      0.08 &      0.31 &      0.11 &      0.03 &       0.02 &        -0.32 &   \bf 6m & 0.53 &      0.21 &      -0.09 &      0.17 &       0.26 &        -0.26 \\

\bf 9m &      0.09 &      0.23 &      0.05 &      -0.01 &       -0.15 &         0.00 &     \bf 9m & 0.54 &      -0.02 &      -0.49 &      -0.31 &       0.02 &         0.02 \\

\bf 12m &         0.09 &          0.15 &         0.04 &         0.01 &         -0.11 &          0.11 &    \bf 12m & 1.11 &          0.56 &         -0.22 &         -0.06 &         -0.02 &          0.11 \\
\midrule
\multicolumn{14}{c}{$\mathbf{M_1}$}  \\
\midrule
\bf 1m &      0.00 &       0.00 &       0.07 &       0.26 &        1.55** &          1.68 &  \bf 1m & 1.81* &       1.65 &       1.81 &       1.40 &         2.59 &          2.28 \\

\bf 3m &      0.00 &      0.00 &       0.03 &       0.24 &        1.45** &         2.66* &  \bf 3m & 2.63** &      2.48* &       2.44** &       2.47** &        2.78** &         3.41** \\

\bf 6m &      0.00 &      0.00 &      0.01* &      0.21* &      1.09** &       2.44** &    \bf 6m & 3.26*** &      3.06*** &      2.53*** &      2.47*** &      2.42*** &       3.02*** \\

\bf 9m &      0.00 &      0.00 &      0.00 &      0.09* &       0.76** &       2.53*** &     \bf 9m & 3.47*** &      3.22*** &      2.81*** &      2.73*** &       2.65*** &       3.19*** \\

\bf 12m &         0.00 &         0.00 &        0.00 &         0.03* &         0.56** &         2.22*** &  \bf 12m & 2.98*** &         3.09*** &        2.66*** &         2.65*** &         2.45*** &         2.84*** \\
\midrule
\multicolumn{14}{c}{$\mathbf{M_2}$} \\
\midrule
\bf 1m &          0.00 &          0.00 &         0.10 &          0.31 &          1.74** &         1.13 &       \bf 1m & 1.74* &          1.52 &         1.67 &         1.37 &          2.54* &          1.70 \\

\bf 3m &          0.00 &          0.00 &         0.04 &         0.31* &          1.39** &         2.32 &      \bf 3m & 1.99* &          1.82 &         1.85* &          1.96* &          2.32** &         2.81* \\

\bf 6m &         0.00 &         0.00 &        0.01 &       0.18* &       1.02** &       2.13** &   \bf 6m & 2.72*** &         2.45*** &        1.93*** &       1.93** &       1.90** &       2.48** \\

\bf 9m &          0.00 &         0.00 &        0.00 &       0.07* &       0.73*** &      2.20** &      \bf 9m & 2.94*** &         2.63*** &        2.21*** &       2.17*** &       2.13*** &      2.64** \\

\bf 12m &        0.00 &        0.00 &        0.00 &         0.03 &     0.60** &        2.00*** &    \bf 12m & 2.56*** &        2.61*** &        2.16*** &         2.19*** &     2.02*** &        2.42*** \\
\midrule
\multicolumn{14}{c}{$\mathbf{M_3}$} \\
\midrule
\bf 1m &      -0.05 &         -0.09 &        -0.61 &         -1.11 &          0.40 &          -2.57 &     \bf 1m & -2.38 &         -3.51 &        -2.28 &         -2.03 &          0.83 &          -2.73 \\

\bf 3m &         -0.03 &         -0.14 &         -0.60 &         -0.95 &         -0.57 &          -2.75 &        \bf 3m & -1.89 &         -2.28 &         -1.60 &         -1.17 &         -0.36 &          -2.70 \\

\bf 6m &        -0.01 &         -0.08 &          -0.43 &        -0.77 &       -0.79 &      -1.52 &    \bf 6m & -1.50 &         -1.89 &          -1.78 &        -1.28 &       -0.63 &      -1.52 \\

\bf 9m &        0.00 &        -0.08 &         -0.39 &      -0.88 &      -1.37&      -1.56 &    \bf 9m & -1.12 &        -1.68 &         -1.96 &      -1.75 &      -1.28 &      -1.55 \\

\bf 12m &       0.00 &        -0.03  &       -0.27 &        -0.69 &         -1.34 &      -1.47 &  \bf 12m & -0.33 &        -0.87 &       -1.52 &        -1.36 &         -1.32 &      -1.46 \\
\midrule
\multicolumn{14}{c}{$\mathbf{M_4}$} \\
\midrule
\bf 1m &         0.08 &          0.19 &          0.27 &          0.35 &         1.88** &          1.87 &      \bf 1m & 1.89 &          1.61 &          1.82 &          1.34 &        2.65* &          2.20 \\

\bf 3m &         0.01 &        0.09* &        0.36** &     0.64* &         1.56* &          2.50 &   \bf 3m & 2.72** &        2.39* &         2.37** &     2.37* &         2.65** &          3.00* \\

\bf 6m &        0.00 &         0.05 &         0.27* &         0.59** &      1.34** &      2.53** &   \bf 6m & 3.52*** &         3.24*** &          2.67*** &         2.59*** &      2.48** &      2.93** \\

\bf 9m &        0.00 &         0.03 &          0.15* &       0.43** &      1.12** &      2.71** &    \bf 9m & 3.65*** &         3.32*** &          2.83*** &       2.75*** &      2.68*** &      3.15*** \\

\bf 12m &        0.00 &        0.02 &        0.07* &         0.26* &        0.89** &      2.43***  &       \bf 12m & 3.28*** &        3.18*** &        2.62*** &         2.62*** &        2.41*** &      2.80*** \\
\midrule
\multicolumn{14}{c}{$\mathbf{M_5}$} \\
\midrule
\bf 1m &         0.00 &       0.02  &         0.20* &          0.45 &       1.81** &          2.36 &    \bf 1m & 1.68* &       1.39 &         1.50 &          1.28 &          2.60* &          2.48 \\

\bf 3m &        0.00 &        0.00 &          0.11* &     0.40* &          1.40** &          2.61* &    \bf 3m & 2.45** &        2.43* &          2.32** &     2.37** &          2.69** &          3.09* \\

\bf 6m &         0.00 &          0.00 &         0.02* &          0.28* &         1.10** &        2.44** &         \bf 6m & 3.42*** &          3.14*** &         2.53*** &          2.42*** &         2.29** & 2.87*** \\

\bf 9m &        0.00 &         0.00 &         0.01 &          0.15* &         0.82** &         2.52*** &         \bf 9m & 3.42*** &         3.13*** &         2.70*** &          2.61*** &         2.54*** &         3.00*** \\

\bf 12m &        0.00 &        0.00 &          0.00 &        0.07* &          0.63** &         2.25*** &         \bf 12m & 3.08*** &        3.09*** &          2.62*** &        2.59*** &          2.38*** &         2.72*** \\
\midrule
\multicolumn{14}{c}{$\mathbf{M_6}$} \\
\midrule
\bf 1m &      0.00 &       0.01 &       0.09 &       0.28 &         1.79** &          2.13 &    \bf 1m & 2.38** &       2.22* &       2.29* &       1.85 &         2.99* &          2.46* \\

\bf 3m &      0.00 &      0.00 &       0.03 &       0.30* &         1.38** &          2.66* &    \bf 3m & 2.71** &      2.57** &       2.46** &       2.48** &         2.73** &          3.22* \\

\bf 6m &      0.00 &      0.00 &      0.01 &      0.18* &       0.97** &         2.31** &    \bf 6m & 3.15*** &      2.92*** &      2.41*** &      2.36*** &       2.28*** &         2.79*** \\

\bf 9m &      0.00 &      0.00 &       0.00 &    0.08* & 0.73** &        2.44*** &     \bf 9m & 3.26*** &      3.04*** &      2.65*** &      2.58*** &       2.50*** &        2.99*** \\

\bf 12m &         0.00 &         0.00 &         0.00 &         0.03* &         0.57** &         2.17*** &          \bf 12m & 2.87*** &         2.97*** &         2.54*** &         2.55*** &         2.36*** &         2.77*** \\
\bottomrule
\end{tabular}
\end{center}
\noindent{\caption*{\scriptsize This table reports annualised certainty equivalent returns (CERs) across alternative models, at different prediction horizons, of $h$= 1-month, 3-month, 6-month, 9-month and 12-month. The coefficient of risk aversion is $\gamma=5$. Panel A presents CERs under the scenario where investors are prevented from extreme investments and as such, portfolio weights are restricted to range in the interval $[-1, 2]$, thus imposing maximum short-selling and leveraging of 100\% respectively. Panel B, reports CER values under the scenario where no allocation restrictions are imposed to investors and, as such, portfolio weights are unbounded, thus allowing for maximum leveraging and short-selling. CERs are generated by out-of-sample forecasts of bond excess returns and are reported in \%. At every time step, $t$, an investor with power utility preferences, evaluates the entire predictive density of bond excess returns and solves the asset allocation problem, thus optimally allocating her wealth between a riskless bond and risky bonds with maturities 2, 3, 4, 5, 7 and 10-years. CER is, then, defined as the value that equates the average utility of each alternative model against the average utility of the EH benchmark. The seven forecasting models used are ATSM with alternative risk price restrictions. Positive values indicate that the models perform better than the EH benchmark. Statistical significance is measured using a one-sided Diebold-Mariano statistic  computed with Newey-West standard errors. * denotes significance at 10\%, ** significance at 5\% and *** significance at 1\% level. The in-sample period is January 1990 to end of 2007, and the out-of-sample period starts in January 2008 and ends at the end of 2018.}}
\end{table}

\begin{table}[!htbp]\scriptsize 
\caption{Out-of-sample Economic performance of Bond excess return forecasts across prediction horizons - Multivariate Asset Allocation}
\label{table:JOINT}
\begin{center}
\begin{tabular}{llllllllllll}
\midrule \midrule
\multicolumn{6}{c}{\bf Panel A: Period - 1985 - 2007} & \multicolumn{6}{c}{\bf Panel B: Period - 1990 - 2018} \\
\midrule
\multicolumn{1}{m{0.5cm}}{} &\multicolumn{1}{m{1cm}}{\bf 1m} &\multicolumn{1}{m{1cm}}{\bf 3m} & \multicolumn{1}{m{1cm}}{\bf 6m} & \multicolumn{1}{m{1cm}}{\bf 9m} & \multicolumn{1}{m{1cm}}{\bf 12m} &
\multicolumn{1}{m{0.5cm}}{} &
\multicolumn{1}{m{1cm}}{\bf 1m} &\multicolumn{1}{m{1cm}}{\bf 3m} & \multicolumn{1}{m{1cm}}{\bf 6m} & \multicolumn{1}{m{1cm}}{\bf 9m} & \multicolumn{1}{m{1cm}}{\bf 12m} \\ 
\midrule
$\mathbf{M_{0}}$ &    -4.35 &    -4.22 &     -3.43 &    -3.43 &  -3.40  &  $\mathbf{M_{0}}$ &  -7.09 &    -2.86 &     -0.24 &    0.61 &  0.57\\

$\mathbf{M_{1}}$ &   0.48 &    0.13 &     2.05** &    2.14** &  1.59** & $\mathbf{M_{1}}$ &   -0.94 &    1.44 &     2.12** &    3.61*** &  3.20*** \\

$\mathbf{M_{2}}$ &   0.63 &    0.40 &     1.84** &    1.84** &  1.36*  & $\mathbf{M_{2}}$ &   -1.26 &    0.89 &     1.65* &    3.03*** &  2.70*** \\

$\mathbf{M_{3}}$ &      0.77 &      0.67 &       1.21** &   0.82* &      0.47   & $\mathbf{M_{3}}$ &   -3.46 &       -3.18 &       -1.47 &   -0.71 &      -0.91  \\

$\mathbf{M_{4}}$ &    -0.76 &       0.11 &       1.25 &   1.26* &      0.79   &  $\mathbf{M_{4}}$ &  2.56 &       1.57 &       2.33** &   3.60*** &      3.13***  \\

$\mathbf{M_{5}}$ &      0.61 &   -0.12 &       0.97 &   0.82 &      0.43   & $\mathbf{M_{5}}$ &  0.67 &   1.08 &       2.11** &   3.45*** &      3.07*** \\

$\mathbf{M_{6}}$ &     1.67 &       0.19 &      1.85** &   1.87** &  1.35**  &  $\mathbf{M_{6}}$ &  -0.06 &       0.84 &      1.91* &   3.37*** &  3.10***  \\
\midrule \midrule
\end{tabular}
\end{center}
\noindent{\caption*{\scriptsize This table reports annualised CERs across alternative models, at different prediction horizons, of $h$= 1-month, 3-month, 6-month, 9-month and 12-month. The coefficient of risk aversion is $\gamma=5$. CERs are generated by out-of-sample forecasts of bond excess returns and are reported in \%. At every time step, $t$, an investor with power utility preferences, evaluates the entire predictive density of bond excess returns and solves the asset allocation problem, thus optimally allocating her wealth across bonds with maturities 2, 3, 4, 5, 7 and 10-years. The seven forecasting models used are DTSMs with alternative risk price restrictions. Positive values indicate that the models perform better than the EH benchmark. Portfolio weights are restricted to range in the interval $[-1, 2]$, thus imposing maximum short-selling and leveraging of 100\% respectively. Panels A and B cover the periods (1985 - 2007) and (1990 - 2018) respectively. Statistical significance is measured using a one-sided Diebold-Mariano statistic computed with Newey-West standard errors.
* denotes significance at 10\%, ** significance at 5\% and *** significance at 1\% level.
}}
\end{table}

\begin{table}[!htbp]\scriptsize
\caption{Out-of-sample Economic performance of Bond excess return forecasts across prediction horizons - Predictive Regressions
}
\label{table:CERSSVSPREDREG}
\begin{center}
\begin{tabular}{lllllllllllllll}
\midrule \midrule 
\multicolumn{14}{c}{\bf Panel A: Period - 1985 - 2007} \\
\midrule                                             \multicolumn{7}{c}{\bf Panel A.1: w= $[-1, 2]$} &         \multicolumn{7}{c}{\bf Panel A.2: w= $[-\infty,+\infty]$} \\
\midrule
\multicolumn{1}{m{0.5cm}}{\bf h} &\multicolumn{1}{m{0.5cm}}{\bf 2Y} &\multicolumn{1}{m{0.5cm}}{\bf 3Y} & \multicolumn{1}{m{0.5cm}}{\bf 4Y} & \multicolumn{1}{m{0.5cm}}{\bf 5Y} & \multicolumn{1}{m{0.5cm}}{\bf 7Y} & \multicolumn{ 1}{m{1cm}}{\bf 10Y} & 
\multicolumn{1}{m{0.5cm}}{\bf h} &\multicolumn{1}{m{0.9cm}}{\bf 2Y} &\multicolumn{1}{m{0.9cm}}{\bf 3Y} & \multicolumn{1}{m{0.9cm}}{\bf 4Y} & \multicolumn{1}{m{0.9cm}}{\bf 5Y} & \multicolumn{1}{m{0.9cm}}{\bf 7Y} & \multicolumn{1}{m{0.5cm}}{\bf 10Y}\\
\midrule                                             \multicolumn{14}{c}{\bf SSVS Predictive Regressions combined with VAR dynamics}  \\
\midrule
\bf 1m &      -0.38 &      -1.92 &      -0.25 &      -1.98 &      -1.76 &      -2.40 &         \bf 1m &      -2.36 &      -3.16 &      -0.11 &      -2.15 &      -1.73 &      -2.42 \\

\bf 3m &      -0.06 &      -0.04 &      -0.25 &      -0.47 &       0.65 &      -0.79 &         \bf 3m &      -0.40 &       0.40 &       0.44 &      -0.17 &       0.59 &      -0.79 \\

\bf 6m &       0.24 &       0.05 &       0.21 &       0.54 &       0.46 &       1.23 &         \bf 6m &       1.97 &       1.53 &       1.80 &       1.80 &       0.43 &       1.24 \\

\bf 9m &       0.26 &       0.32 &       0.36 &       0.50 &       1.36 &       1.71 &         \bf 9m &       3.56 &       3.43 &       2.89 &       2.81 &       2.09 &       1.74 \\

\bf 12m &       0.23 &       0.35 &       0.29 &       0.51 &       1.12 &       1.87 &        \bf 12m &       3.72 &      4.58* &       3.26 &       3.34 &       2.56 &       2.02 \\
\midrule
\multicolumn{14}{c}{\bf Panel B: Period - 1990 - 2018} \\
\midrule                                             \multicolumn{7}{c}{\bf Panel B.1: w= $[-1,2]$} &         \multicolumn{7}{c}{\bf Panel B.2: w= $[-\infty,+\infty]$} \\
\midrule
\multicolumn{1}{m{0.5cm}}{\bf h} &\multicolumn{1}{m{0.5cm}}{\bf 2Y} &\multicolumn{1}{m{0.5cm}}{\bf 3Y} & \multicolumn{1}{m{0.5cm}}{\bf 4Y} & \multicolumn{1}{m{0.5cm}}{\bf 5Y} & \multicolumn{1}{m{0.5cm}}{\bf 7Y} & 
\multicolumn{ 1}{m{1cm}}{\bf 10Y} & 
\multicolumn{1}{m{0.5cm}}{\bf h} &\multicolumn{1}{m{0.9cm}}{\bf 2Y} &\multicolumn{1}{m{0.9cm}}{\bf 3Y} & \multicolumn{1}{m{0.9cm}}{\bf 4Y} & \multicolumn{1}{m{0.9cm}}{\bf 5Y} & \multicolumn{1}{m{0.9cm}}{\bf 7Y} & \multicolumn{1}{m{0.5cm}}{\bf 10Y}\\
\midrule                                             \multicolumn{14}{c}{\bf SSVS Predictive Regressions combined with VAR dynamics}  \\
\midrule
\bf 1m &      -0.24 &      -0.55 &      -0.35 &      -2.17 &      -0.59 &       1.17 &         \bf 1m &      -4.03 &      -4.27 &      -0.65 &      -3.13 &      -0.17 &      1.68* \\

\bf 3m &       0.42 &       0.61 &       0.39 &       0.75 &       0.98 &       1.82 &         \bf 3m &       2.14 &       2.96 &       2.37 &       2.34 &      -0.01 &       2.22 \\

\bf 6m &       0.54 &       0.75 &       0.80 &       0.86 &       1.53 &       2.56 &         \bf 6m &       2.52 &      3.45* &       2.77 &       2.55 &       2.45 &       2.47 \\

\bf 9m &     0.55** &     0.97** &      1.30* &       1.55 &       1.98 &       2.97 &         \bf 9m &      1.92* &     4.08** &     4.34** &      4.52* &       3.71 &       3.28 \\

\bf 12m &    0.54*** &    1.07*** &     1.45** &       1.73 &       2.19 &       3.12 &        \bf 12m &     2.99** &    5.29*** &    5.08*** &     6.18** &       4.08 &       3.51 \\
\midrule \midrule 
\end{tabular}
\end{center}
\noindent{\caption*{\scriptsize This table reports annualised CERs, at different prediction horizons, of $h$= (1,3,5,9,12)-month. The coefficient of risk aversion is $\gamma=5$. CERs are generated by out-of-sample forecasts of bond excess returns and are reported in \%. At every time step, $t$, an investor with power utility preferences, evaluates the entire predictive density of bond excess returns and solves the asset allocation problem, thus optimally allocating her wealth between a riskless bond and risky bonds with maturities 2, 3, 4, 5, 7 and 10-years. 
Positive values indicate that the models perform better than the EH benchmark. Panels A.1 and B.1 present CERs under the  scenario of portfolio weights being restricted to range in the interval $[-1, 2]$, thus imposing maximum short-selling and leveraging of 100\% respectively. Panels A.2 and B.2, report CER values under the second scenario, where no allocation restrictions are imposed to investors. Panel A and B cover the periods (1985 - 2007) and (1990 - 2018) respectively. Statistical significance is measured using a one-sided Diebold-Mariano statistic  computed with Newey-West standard errors. * denotes significance at 10\%, ** significance at 5\% and *** significance at 1\% level. 
}}
\end{table}

\begin{appendices}

\newpage

\title{Online Appendix: Sequential Learning and Economic Benefits from Dynamic Term Structure Models}

\section*{Appendix A: Model Specification Details}

Given $P_{t}^{n+1}=E_{t} \left ( M_{t+1}P_{t+1}^{n} \right )$, it follows that bond prices are exponentially affine functions of the state vector (see, \cite{Duffie96})
\begin{equation}
\label{bond price}
P_{t}^{n} = \exp(A_{n} + B_{n}' X_{t}),\;\;n=1,\dots,J
\end{equation}
with $A_{n}$ being a scalar and $B_{n}$ a $N \times 1$ vector satisfying the following recursions
\begin{align}
\label{recursions} 
A_{n+1} &= A_{n} + B_{n}' ( \mu - \lambda_{0} ) + \frac{1}{2} B_{n}' \Sigma \Sigma' B_{n} - \delta_{0} \\ 
B_{m+1} &=  B_{n}'( \Phi - \lambda_{1} ) - \delta_{1}
\end{align}
with $A_{0} = 0$ and $B_{0} = 0$, leading to the $\mathbb{Q}$ dynamics of equation (2) of the main paper
$$
X_{t}-X_{t-1}= \mu^{\mathbb{Q}} + \Phi^{\mathbb{Q}} X_{t-1} + \Sigma \varepsilon_{t}^{\mathbb{Q}}
$$
To derive expressions for the  parameters in equation (6) and (7) in the main body of the paper, we apply the transformation of (5) to the model of (1) as shown below:
\label{appendix:modeldetails}
\begin{align}
\label{VAR_P_Q_details}
\mu_{\mathcal{P}}^{\mathbb{Q}} &=W B_{n,x} \mu^{\mathbb{Q}} - \Phi_{\mathcal{P}}^{\mathbb{Q}}W A_{n,X}\\
\Phi_{\mathcal{P}}^{\mathbb{Q}} &= W B_{n,x} \Phi^{\mathbb{Q}} (W B_{n,x})^{-1}\\
\Sigma_{\mathcal{P}} &= (W B_{n,x}) \Sigma
\end{align}

In a similar manner we can also obtain the matrices in the Ricatti recursions:

\begin{align}\label{rotation2} 
A_{n,\mathcal{P}} &= A_{n,X} - B_{n,X} ( W B_{n,X} )^{-1}( W A_{n,X} ) \\ 
B_{n,\mathcal{P}} &=  B_{n,X} ( W B_{n,X} )^{-1}
\end{align}


\section*{Appendix B: Prior Specification and Algorithmic Details}

\subsection*{B.1: Prior Specification}

As mentioned in the main body of the paper, the prior on the elements of the $\lambda^{\mathcal{P}}$ are specified by the spike and slab formulation 
\begin{equation}\label{spikeslab_app}
\lambda^{\mathcal{P}}_{ij}\sim (1-\gamma_{ij})N(0,\tau_{ij}^{(0)})+\gamma_{ij} N(0,\tau_{ij}^{(1)}),  
\end{equation}
where the slab variances are specified by the Zellner g-prior, where the variance takes the form of $g \hat{V}$. For the calculation of $\hat{V}$, we follow the procedure in the appendix C.3 of \cite{Bauer18} and set $g=\mbox{max}(T,p^2)$ ($p$ denoting the number of parameters), which in our case reduces to $T$ that is always larger.

In order to explore the robustness of the results to the choice of $g$ we implemented versions of the corresponding sequential SSVS algorithms, based on $g=T$ and $g=p^2$, in the period (1990-2018) and calculated annualised CERs, across different maturities and prediction horizons. Portfolio weights constrained to $[-1,2]$. Table \ref{table:SSVSB1205} contains the relevant results, which are quite similar, thus suggesting a minimal impact of $g$ as long as it is set to a reasonable value.

\begin{table}[!ht]\scriptsize \caption{Out-of-sample economic performance of bond excess return forecasts for model $M_4$ across different prior specifications with portfolio weights restricted to $[-1,2]$: multiple prediction horizons and maturities - Period: 1990-2018.}
\label{table:SSVSB1205}
\begin{center}
\begin{tabular}{llllllll}
\toprule
\multicolumn{1}{m{1cm}}{\bf h} &\multicolumn{1}{m{0.5cm}}{\bf 2Y} &\multicolumn{1}{m{0.5cm}}{\bf 3Y} & \multicolumn{1}{m{0.5cm}}{\bf 4Y} & \multicolumn{1}{m{0.5cm}}{\bf 5Y} & \multicolumn{1}{m{0.5cm}}{\bf 7Y} & \multicolumn{1}{m{1cm}}{\bf 10Y}& \multicolumn{1}{m{1cm}}{\bf Joint}\\
\midrule 
\multicolumn{8}{c}{\bf $g=T$}  \\
\midrule
\bf 1m & 0.08 &       0.19 &       0.27 &       0.35 &     1.88** &       1.87 &       2.56 \\
\bf 3m & 0.01 &      0.09* &     0.36** &      0.64* &      1.56* &       2.50 &       1.57 \\
\bf 6m & 0.00 &       0.05 &      0.27* &     0.59** &     1.34** &     2.53** &     2.33** \\
\bf 9m & 0.00 &       0.03 &      0.15* &     0.43** &     1.12** &     2.71** &    3.60*** \\
\bf 12m & 0.00 &       0.02 &      0.07* &      0.26* &     0.89** &    2.43*** &    3.13*** \\
\midrule 
\multicolumn{8}{c}{\bf $g=p^2$}  \\
\midrule
\bf 1m & 0.03 &      0.16* &       0.27 &       0.35 &      1.55* &       0.75 &       0.55 \\
\bf 3m & 0.04 &      0.20* &     0.49** &      0.72* &      1.63* &       2.51 &       1.49 \\
\bf 6m & 0.04 &      0.16* &      0.41* &      0.71* &      1.38* &     2.49** &      2.12* \\
\bf 9m & 0.02 &       0.09 &      0.28* &     0.59** &     1.22** &     2.72** &    3.42*** \\
\bf 12m & 0.01 &       0.04 &      0.17* &      0.40* &     0.97** &    2.41*** &    3.03*** \\
\bottomrule
\end{tabular}
\end{center}
\end{table}

For the remaining parameters, for which we are not seeking restrictions and have therefore no concerns on issues such as the Lindley paradox, low informative priors were assigned, although some relevant information is available in this context (see, \cite{Chib09}). We first transform all restricted range parameters so that they have unrestricted range and scale parameters which typically take very small values. 
Specifically, we consider a Cholesky factorization of $\Sigma_{\mathcal{P}}$ where the diagonal elements are transformed to the real line and off-diagonal elements are scaled by $10^6$. In order to preserve the ordering of the eigenvalues $g^{\mathbb{Q}}$ we apply a reparametrisation and work with their increments that are again transformed to the real line. Finally, we scale $k_{\infty}^{\mathbb{Q}}$ by $10^6$ as well.
Independent Normal distributions with zero means and large variances are then assigned to each transformed or scaled parameter. Finally, for $\sigma_e^2$, we assign the conjugate Inverse-Gamma prior as in \cite{Bauer18}.

\subsection*{B.2: Information on Sequential Monte Carlo Implementation}

The IBIS algorithm was implemented with the following specifications. The number of particles $N_{\theta}$ was set to $2000$, see subsection B.3 for some justification on this choice. The resampling and jittering steps were triggered in cases where the effective sample size (ESS) was lower than $1400$. In order to ensure good quality, in the sense of Monte Carlo error, in the posterior and predictive distribution estimates, we used adaptive tempering such that the ESS is always at least $1400$.  Adaptive tempering serves the purpose of smoothing peaked likelihoods. It is achieved by bridging two successive targets via an intermediate target sequence. The idea is to modify the sequence of target distributions so that it evolves from the prior to the posterior more smoothly (see \cite{jasra2011} and \cite{Kantas2014} for details). Implementation of the IBIS algorithm with hybrid adaptive tempering steps is outlined in Table \ref{tab:ibisat}.

\begin{table}[!ht]
\begin{flushleft}
\hrule
\medskip
{\itshape \small
\vspace{0.2cm}
Initialize $N_{\theta}$ particles by drawing independently $\theta_{i}\sim \pi(\theta)$ with importance weights $\omega_{i}=1$, $i=1,\dots,N_{\theta}$. For $t,\dots,T$ and each time for all $i$:\vspace{0.2cm}
\begin{itemize}
\item[1] Set $\omega_{i}'=\omega_{i}$. 
\item[2] Calculate the incremental weights 
$u_t(\theta_{i})=f\big(Y_{t}|Y_{t-1},\theta_{i})$.
\item[3] Update the importance weights $\omega_{i}$ to $\omega_{i}u_t(\theta_{i})$.
\item[4] If degeneracy criterion ESS($\omega$) is triggered, perform the following sub-steps:
\begin{itemize}
\item[(a)] Set $\phi=0$ and $\phi'=0$.
\item[(b)] While $\phi<1$
\begin{itemize}
\item[i.] If degeneracy criterion ESS($\omega''$) is not triggered, where $\omega_{i}''=\omega_{i}'[u_t(\theta_{i})]^{1-\phi'}$, set $\phi=1$, otherwise find $\phi\in[\phi',1]$ such that ESS($\omega'''$) is greater than or equal to the trigger, where $\omega_{i}'''=\omega_{i}'[u_t(\theta_{i})]^{\phi-\phi'}$, for example using bisection method, see \cite{Kantas2014}.
\item[ii.] Update the importance weights $\omega_{i}$ to $\omega_{i}'[u_t(\theta_{i})]^{\phi-\phi'}$.
\item[iii.] Resample: Sample with replacement $N_{\theta}$ times from the set of $\theta_{i}$s according to their weights $\omega_i$.
\item[iv.] Jitter: Replace $\theta_i$s with $\tilde{\theta}_i$s by running MCMC chains with each $\theta_i$ as input and $\tilde{\theta}_i$ as output, using likelihood given by $f(Y_{0:t-1}|\theta_{i})[f\big(Y_{t}|\theta_{i})]^{\phi}$. 
\item[v.] Calculate the incremental weights $u_t(\theta_{i})=f\big(Y_{t}|Y_{t-1},\theta_{i})$.
\item[vi.] Set $\omega_{i}'=\omega_{i}$ and $\phi'=\phi$.
\end{itemize}
\end{itemize} 
\end{itemize}}
\medskip
\hrule
\end{flushleft}
\caption{IBIS algorithm with hybrid adaptive tempering}
\label{tab:ibisat}
\end{table}

For the jittering step, the MCMC sampler used was very similar to \cite{Bauer18}. This algorithm adopts a Gibbs scheme splitting the parameters into the blocks consisting of $\sigma_e^2$, $\gamma_{ij}$s, $\lambda_{0\mathcal{P}}$ and $\lambda_{1\mathcal{P}}$, $\Sigma_{\mathcal{P}}$, and the $\mathbb{Q}$ parameters. For the first three blocks the full conditional posteriors can be identified, whereas for $\Sigma_{\mathcal{P}}$, and the $\mathbb{Q}$ parameters independence samplers may be constructed, using the maximum likelihood estimate as the mean in respective proposal density and negative inverse of the corresponding Hessian as its covariance. In some cases we deviated from \cite{Bauer18} and obtained the proposal distributions of the independence samplers for updating $\Sigma_{\mathcal{P}}$ and the $\mathbb{Q}$ parameters, based on estimates of their mean and variance from the IBIS output, thus taking advantage of synergies between the two approaches. The MCMC algorithm is generally quite efficient in terms of mixing and therefore can quickly drift off from its initial values. For this reason we observed that a low number of MCMC iterations per particle in the jittering step is sufficient; we ended up using $5$ iterations for each particle. It is good practice to monitor the acceptance rate of the independence samples across particles as well as correlations of particles for each parameter before and after jittering. In our case these were quite reasonable with acceptance rates being at least 40\% and most correlations being at most 0.2.

\subsection*{B.3: Monte Carlo Error and Number of Particles}

As the IBIS algorithm is used to generate Monte Carlo estimates of the economic value quantities such as certainty equivalent returns (CER) and utilities, it is important to monitor the Monte Carlo error in these estimates and ensure it is small compared to the other sources of variability. A user-specified parameter that can be used to reduce the amount of Monte Carlo error is the number of particles in the IBIS algorithm. It is also important to note that in the tables with indices on economic value, each cell provides the CER, which is mainly determined by the ratio of the cumulative utility of the model in question against the benchmark model. However, the stars that indicate the significance are determined by Diebold-Mariano test on the difference between the utilities of these two models. In order to assess the impact of the Monte Carlo error we focus on the variance of the difference between these utilities and in particular the amount of which can be attributed to Monte Carlo error.

To explore the above we ran $20$ independent instances of the IBIS algorithm using the $M_1$ model on data from the time period between 1985 and the end of 2007, where the out-of-sample period is from 1997. The process was repeated for all prediction horizons and across maturities, whereas we also considered $1000$, $2000$ and $5000$ as choices regarding the number of particles. In each of these combinations the variance of the Monte Carlo estimator, of utility differences, can be estimated as the sample variance of the Monte Carlo estimates taken over these $20$ samples. On the other hand, the overall variance can be estimated as the sample mean of the variances, calculated at each run, again taken over these $20$ samples. Table \ref{table:M1postest19902020} shows the percentage or the Monte Carlo variance against the total variance for each of the previously mentioned combinations. 

\begin{table}[!ht]\scriptsize \caption{Monte Carlo errors in \% for $M_1$ model across 1000, 2000 and 5000 particles - Period: January 1985 - end of 2007.}
\label{table:M1postest19902020}
\begin{center}
\begin{tabular}{r|rrrrrr}
\toprule
     $\bf h  \backslash  n$ &         \bf 2Y &         \bf 3Y &         \bf 4Y &         \bf 5Y &         \bf 7Y &        \bf 10Y \\
\midrule
\bf \% & \multicolumn{6}{c}{1000 particles} \\
\midrule
         \bf 1m &       0.19 &       0.27 &       0.19 &       0.13 &       0.13 &       0.31 \\
         \bf 3m &       0.13 &       0.18 &       0.16 &       0.11 &       0.15 &       0.09 \\
         \bf 6m &       0.08 &       0.10 &       0.06 &       0.06 &       0.06 &       0.07 \\
         \bf 9m &       0.08 &       0.09 &       0.09 &       0.11 &       0.04 &       0.06 \\
        \bf 12m &       0.06 &       0.06 &       0.07 &       0.03 &       0.05 &       0.07 \\
\midrule
\bf \% & \multicolumn{6}{c}{2000 particles} \\
\midrule
         \bf 1m &       0.07 &       0.09 &       0.08 &       0.07 &       0.10 &       0.08 \\
         \bf 3m &       0.03 &       0.05 &       0.03 &       0.03 &       0.05 &       0.07 \\
         \bf 6m &       0.07 &       0.06 &       0.03 &       0.06 &       0.03 &       0.03 \\
         \bf 9m &       0.05 &       0.06 &       0.04 &       0.05 &       0.05 &       0.02 \\
        \bf 12m &       0.05 &       0.07 &       0.08 &       0.04 &       0.03 &       0.02 \\
\midrule
\bf \% & \multicolumn{6}{c}{5000 particles} \\
\midrule
         \bf 1m &       0.04 &       0.04 &       0.05 &       0.05 &       0.03 &       0.03 \\
         \bf 3m &       0.01 &       0.01 &       0.02 &       0.01 &       0.02 &       0.01 \\
         \bf 6m &       0.02 &       0.01 &       0.02 &       0.01 &       0.01 &       0.02 \\
         \bf 9m &       0.01 &       0.01 &       0.01 &       0.02 &       0.02 &       0.01 \\
        \bf 12m &       0.03 &       0.01 &       0.01 &       0.01 &       0.02 &       0.01 \\
\bottomrule
\end{tabular}
\end{center}
\noindent{\caption*{\scriptsize Out-of-sample period is from 1997.}}
\end{table}

As shown in this table, the Monte Carlo variance is at most $0.31$\% of the total variance and this refers to the case of $1000$ particles. This value is quite low and offers reassurance that the Monte Carlo error is not substantial. Increasing to $2000$ particles seems to more or less halve these percentages that are now below $0.1$\%. A further increase to $5000$ particles reduced these percentages even more, as expected. Nevertheless, since these numbers were already quite low we decided to set the number of particles to $2000$.

\newpage

\section*{Appendix C: Information on Data and extraction of Principal Components}

\begin{figure}[!htbp]\scriptsize
\begin{center}
\includegraphics[trim = 25mm 20mm 0mm 0mm, height=4.8in,left]{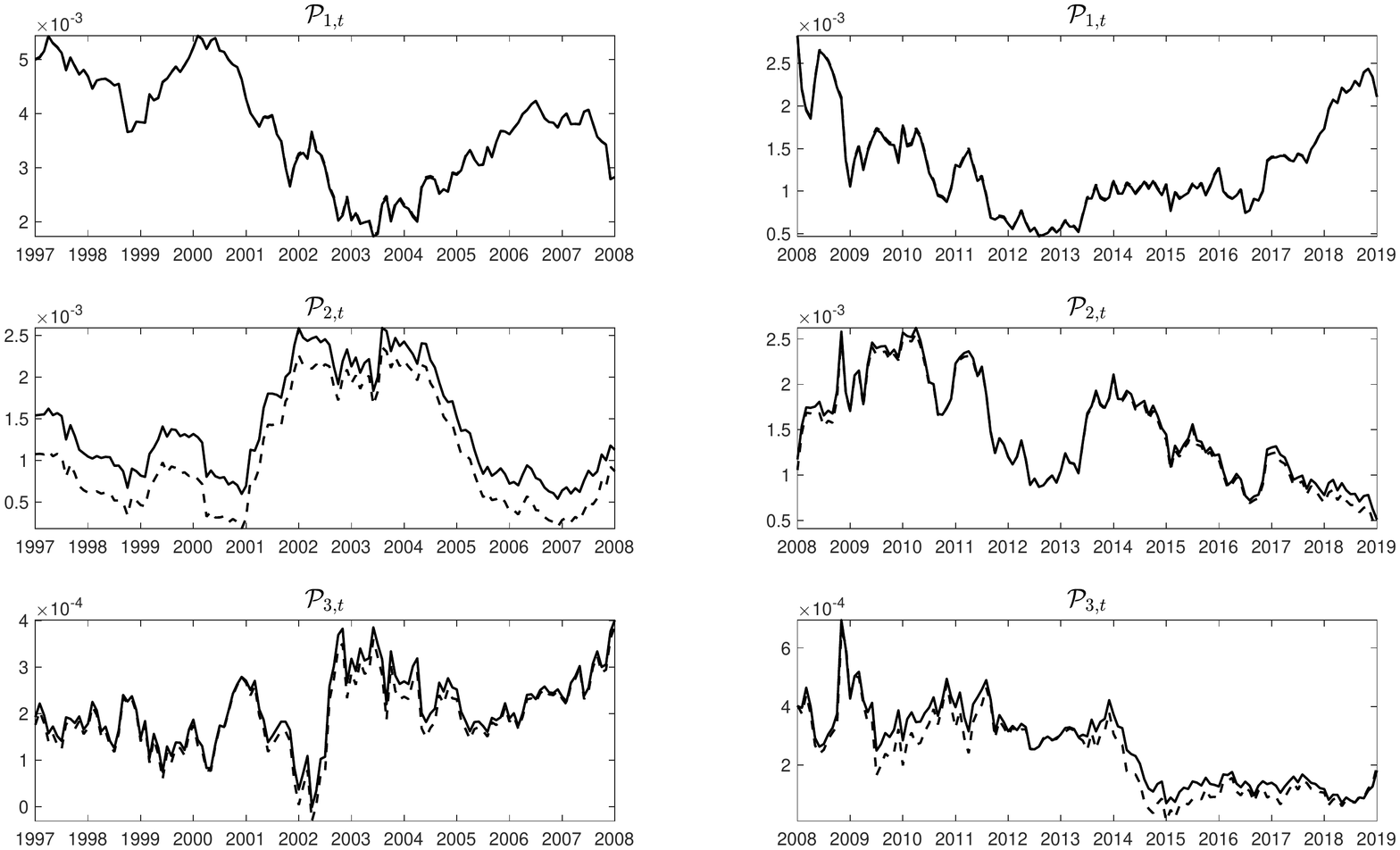}
\caption{Principal Components (PCs) extracted from the yield curve for the two periods where the predictions were evaluated (January $1997$ - end of $2007$ and January $2008$ - end of $2018$). The dashed lines correspond to principal components based on loadings calculated from the entire sub-samples (January $1980$ - end of $2007$ and January $1985$ - end of $2018$). The solid lines, which are the ones used for the data analysis, were based on loadings calculated from the warm-up periods only (January $1980$ - end of $1996$ and January $1985$ - end of $2007$).} 
\label{fig:PCs}
\end{center}
\end{figure}

 Extracting the first three principal components from the yield curve can be done either by calculating the principal component loadings from the warm-up or the entire period; the differences in the resulting principal component time series are depicted in Figure \ref{fig:PCs}. The correlations between these series reach levels above $0.99$, suggesting negligible differences. Nevertheless, in order to prevent any data leaking issues when assessing the predictive performance, the loadings from the warm-up period only are used. 

\newpage

The data also covers the two recession periods of $2001$-$2002$ and $2008$-$2009$, where the yield curve shared qualitatively similar characteristics. Both episodes started from a flat yield curve during the pre-recession periods, following an inversion of the curve that reflects an increase in short rates due to expectations about the Fed tightening its policy, and then a steepening as a reaction to policy adjustments by the Fed reflecting strong growth and inflation expectations. This is reflected in the two principal components, with PC1 decreasing and PC2 increasing during the recession periods.

The data contain several important events and it is interesting to examine the trajectories of the  principal components during these periods, provided by Figure $\ref{fig:PCs}$. More specifically, during the period between $2004$ and $2006$, referred to by former Fed Chairman Greenspan as the `conundrum period', the Federal Reserve applied a tight monetary policy by substantially increasing its target federal funds rate by 4\%, reaching a value of 5.25\% by mid-$2006$. At the same time, long-term yields actually declined, directly affecting the shape of the yield curve, which flattened pronouncedly. This episode is captured by the first two implied PCs, as shown in Figure \ref{fig:PCs}. In particular, PC1 increased substantially during the aforementioned period, reflecting the increase in the level of the term structure, while PC2 decreased to very low levels, reflecting the flattening of the curve.


\section*{Appendix D: Assessing Predictive Performance via log predictive score}

The out-of-sample $R^2$ measure, used to assess the predictive ability of the models, only focused on a point summary of the predictive distribution and may ignore important information. To assess the entire predictive distribution offered by our scheme, rather than just point predictions, we use the log score (LS); a standard choice among scoring rules with several desirable properties, such as being strictly proper, see for example \cite{Dawid14}. The log score of the predictive distribution can be approximated numerically using kernel methods. For the EH case, this evaluation can be done analytically. 
Similarly to out-of-sample $R^2$, LS is aggregated over all prediction times ($t_0$ to $t$) and maturities. In order to get a feeling for how large the differences from the EH benchmark are, we report the p-values from the Diebold-Mariano test (see, \cite{Gargano19}) noting that these are viewed as indices rather than formal hypothesis tests. They are based on t-statistics computed taking into account potential serial correlations in the standard errors (see, \cite{NeweyWest1987}).  

 
 \begin{table}[!htbp]\scriptsize						\caption{Out-of-sample statistical performance of Bond excess return forecasts measured via log predictive score - Multiple prediction horizons - Period: January 1985 - end of 2007.}
\label{table:LS2007}
\begin{center}
\begin{tabular}{lllllll}
\toprule
\multicolumn{1}{m{1.5cm}}{$\bf (h \backslash \bf n-h)$} &\multicolumn{1}{m{1.3cm}}{\bf 2Y} &\multicolumn{1}{m{1.3cm}}{\bf 3Y} & \multicolumn{1}{m{1.3cm}}{\bf 4Y} & \multicolumn{1}{m{1.3cm}}{\bf 5Y} & \multicolumn{1}{m{1.3cm}}{\bf 7Y} & \multicolumn{ 1}{m{1.3cm}}{\bf 10Y} \\
\midrule                                               \multicolumn{7}{c}{$\mathbf{M_0}$} \\
\midrule
\textbf{1m} &    -0.01 &     0.00 &     0.00 &     -0.01 &       -0.02 &    -0.03 \\

\textbf{3m} &    0.05 &    0.04 &    0.04 &    0.01 &     -0.01 &       -0.07 \\

\textbf{6m} &    0.06 &    0.03 &    0.04 &    0.02 &    -0.04 &      -0.09 \\

\textbf{9m} &    0.01 &    0.00 &    0.01 &    0.00 &    -0.04 &    -0.07 \\

\textbf{12m} &    -0.05 &    -0.05 &    -0.03 &    -0.02 &    -0.06 &    -0.08 \\
\midrule                                               \multicolumn{ 7}{c}{$\mathbf{M_1}$} \\
\midrule
        \textbf{1m} &    0.01 &     0.00 &     0.02** &     0.00 &       0.00 &       0.01 \\

        \textbf{3m} &    0.04**  &    0.04** &    0.04*** &    0.03* &     0.02* &       0.01 \\

  \textbf{6m} &    0.05 &    0.05* &    0.06** &    0.05* &    0.04* &     0.03* \\

 \textbf{9m} &    0.03 &    0.05 &    0.06* &    0.05 &    0.04 &       0.05* \\

\textbf{12m} &    0.01 &    0.03 &    0.05 &    0.05 &     0.05 &     0.05* \\
\midrule 
\multicolumn{ 7}{c}{$\mathbf{M_2}$} \\
\midrule 
        \textbf{1m} &   -0.01 &     0.01 &       0.02* &      0.00 &       -0.01 &       -0.01 \\

    \textbf{3m} &    0.04** &     0.04** &      0.05** &       0.02 &       0.02 &       0.01 \\

\textbf{6m} &    0.06** &    0.06** &     0.06** &     0.03* &     0.03 &      0.03 \\

 \textbf{9m} &    0.04 &    0.06 &     0.06* &     0.03 &     0.03 &     0.03 \\

\textbf{12m} &    -0.01 &       0.03 &       0.04 &       0.05 &       0.03 &       0.03 \\
\midrule                                               \multicolumn{ 7}{c}{$\mathbf{M_3}$} \\
\midrule
        \textbf{1m} &    -0.01 &    0.00 &      0.01 &     0.00 &       -0.01 &       0.00 \\

 \textbf{3m} &    0.05** &     0.04* &       0.04** &       0.03** &       0.01 &       0.01 \\

 \textbf{6m} &    0.05 &    0.05 &     0.05** &       0.05*** &       0.03** &       0.03*** \\

\textbf{9m} &    0.06 &     0.07* &     0.08** &       0.06** &       0.05** &       0.04*** \\

\textbf{12m} &    0.05 &       0.06 &       0.08* &      0.07** &       0.06*** &      0.06*** \\
\midrule                                               \multicolumn{ 7}{c}{$\mathbf{M_4}$} \\
\midrule
        \textbf{1m} &    0.00 &     0.01 &      0.00 &     0.00 &       -0.01 &      0.00 \\

 \textbf{3m} &    0.05** &     0.03* &       0.05*** &       0.04*** &       0.03** &      0.00 \\

\textbf{6m} &    0.06* &     0.07** &      0.06*** &       0.04* &       0.05** &      0.04** \\

 \textbf{9m} &    0.05 &     0.06 &      0.07** &       0.06** &       0.05* &       0.04** \\

\textbf{12m} &    0.03 &       0.03 &       0.05 &      0.05 &      0.05 &      0.05* \\
\midrule                                               \multicolumn{ 7}{c}{$\mathbf{M_5}$} \\
\midrule
        \textbf{1m} &     0.00 &       0.00 &       0.01 &       -0.01 &       -0.01 &      0.00 \\

\textbf{3m} &    0.04* &    0.03* &     0.04** &      0.03*** &      0.02* &      0.01 \\

\textbf{6m} &    0.05 &    0.04 &    0.04 &    0.02 &    0.03 &       0.03 \\

\textbf{9m} &    0.03 &    0.04 &    0.05 &    0.05 &    0.04 &      0.03 \\

\textbf{12m} &    0.00 &    0.02 &    0.03 &    0.05 &    0.03 &      0.04* \\
\midrule                                               \multicolumn{ 7}{c}{$\mathbf{M_6}$} \\
\midrule
        \textbf{1m} &    0.01 &     0.02** &     0.01 &     0.00 &       0.00 &       0.00 \\

 \textbf{3m} &    0.04** &    0.04** &    0.04*** &    0.02* &      0.03* &       0.01 \\

 \textbf{6m} &    0.04 &    0.05** &    0.06*** &    0.05** &    0.04** &      0.04** \\

\textbf{9m} &    0.05 &    0.06* &    0.07** &    0.06** &    0.06** &     0.05** \\

\textbf{12m} &    0.02 &    0.04 &    0.06* &    0.06* &     0.05* &     0.05* \\
\bottomrule
\end{tabular}
\end{center}
\noindent{\caption*{\footnotesize This table reports out-of-sample log predictive score (LS) across alternative models, at different prediction horizons, of $h$= 1-month, 3-month, 6-month, 9-month and 12-month. The seven forecasting models used are ATSM with alternative risk price restrictions. 
The EH implies the historical mean being the optimal forecast of excess returns. Positive values of this statistic imply that the forecast outperforms the historical mean forecast and suggests evidence of time-varying return predictability. Statistical significance is measured using a one-sided Diebold-Mariano statistic computed with Newey-West standard errors. * denotes significance at 10\%, ** significance at 5\% and *** significance at 1\% level. The in-sample period is January 1985 to end of 1996, and the out-of-sample period starts in January 1997 and ends in end of 2007.}}
\end{table}

\begin{table}[!htbp]\scriptsize						\caption{Out-of-sample statistical performance of Bond excess return forecasts measured via log predictive score - Multiple prediction horizons - Period: January 1990 - end of 2018.}
\label{table:LS2018}
\begin{center}
\begin{tabular}{lllllll}
\toprule
\multicolumn{1}{m{1.5cm}}{$\bf (h \backslash \bf n-h)$} &\multicolumn{1}{m{1.3cm}}{\bf 2Y} &\multicolumn{1}{m{1.3cm}}{\bf 3Y} & \multicolumn{1}{m{1.3cm}}{\bf 4Y} & \multicolumn{1}{m{1.3cm}}{\bf 5Y} & \multicolumn{1}{m{1.3cm}}{\bf 7Y} & \multicolumn{ 1}{m{1.3cm}}{\bf 10Y} \\
\midrule                                               \multicolumn{7}{c}{\textbf{M0}} \\
\midrule
\textbf{1m} &    -0.01 &     -0.01 &     0.00 &     -0.01 &       0.01 &    -0.05 \\

\textbf{3m} &    0.10*** &    0.05 &    0.03 &    0.01 &     -0.03 &       -0.02 \\

\textbf{6m} &    0.18*** &    0.12** &    0.06 &    0.05 &    0.05 &      0.01 \\

\textbf{9m} &    0.23*** &    0.13** &    0.06 &    0.03 &    0.03 &    0.02 \\

\textbf{12m} &    0.31*** &    0.20** &   0.10 &    0.07 &    0.05 &    0.04 \\
\midrule                                               \multicolumn{ 7}{c}{\textbf{M1}} \\
\midrule
        \textbf{1m} &    0.00 &     0.00 &     -0.01 &     -0.01 &       -0.01 &       -0.03 \\

        \textbf{3m} &    0.10***  &    0.09*** &    0.06*** &    0.05*** &     0.05** &    0.04* \\

  \textbf{6m} &    0.16*** &    0.13*** &    0.09*** &    0.08** &    0.07*** &     0.08*** \\

 \textbf{9m} &    0.21*** &    0.15*** &    0.11*** &    0.09*** &    0.09*** &       0.11*** \\

\textbf{12m} &    0.20*** &    0.18*** &    0.15*** &    0.13*** &     0.12*** &     0.14*** \\
\midrule 
\multicolumn{ 7}{c}{\textbf{M2}} \\
\midrule 
        \textbf{1m} &   0.00 &     -0.01 &       0.00 &      0.01 &       0.00 &       -0.02 \\

    \textbf{3m} &    0.11*** &     0.08*** &      0.06*** &       0.04** &       0.04** &   0.04* \\

\textbf{6m} &    0.18*** &    0.14*** &     0.09*** &     0.08** &     0.07*** &      0.08** \\

 \textbf{9m} &    0.22*** &    0.16*** &     0.12*** &     0.09*** &     0.08** &     0.10*** \\

\textbf{12m} &    0.24*** &       0.21*** &     0.16*** &       0.12*** &       0.12*** &       0.13*** \\
\midrule                                               \multicolumn{ 7}{c}{\textbf{M3}} \\
\midrule
        \textbf{1m} &    0.00 &    -0.01 &     -0.02 &     -0.02 &       -0.01 &       -0.05 \\

 \textbf{3m} &    0.11*** &     0.06** &       0.04* &       0.01 &       0.00 &       -0.03 \\

 \textbf{6m} &    0.17*** &    0.09** &     0.04 &       0.02 &       0.01 &       -0.02 \\

\textbf{9m} &    0.26*** &     0.13** &     0.05 &       0.01 &       0.01 &       -0.03 \\

\textbf{12m} &    0.35*** &       0.19*** &       0.10* &      0.05 &       0.00 &      -0.04 \\
\midrule                                               \multicolumn{ 7}{c}{\textbf{M4}} \\
\midrule
        \textbf{1m} &    0.00 &     -0.01 &      0.00 &     -0.01 &       0.00 &      -0.04 \\

 \textbf{3m} &    0.10*** &     0.08*** &       0.06*** &       0.05*** &       0.05*** &      0.04* \\

\textbf{6m} &    0.17*** &     0.13*** &      0.11*** &       0.09*** &       0.08*** &      0.09** \\

 \textbf{9m} &    0.23*** &     0.16*** &      0.12*** &       0.11*** &       0.10*** &       0.12*** \\

\textbf{12m} &    0.25*** &       0.20*** &       0.16*** &      0.13*** &      0.12*** &      0.14*** \\
\midrule                                               \multicolumn{ 7}{c}{\textbf{M5}} \\
\midrule
        \textbf{1m} &     0.00 &       0.00 &       0.00 &       -0.02 &       -0.02 &      -0.04 \\

\textbf{3m} &    0.11*** &    0.10*** &     0.06*** &      0.04*** &      0.05** &      0.03 \\

\textbf{6m} &    0.17*** &    0.14*** &    0.09*** &    0.08*** &    0.08*** &       0.09** \\

\textbf{9m} &    0.21*** &    0.16*** &    0.11*** &    0.10*** &    0.10*** &      0.11*** \\

\textbf{12m} &    0.21*** &    0.18*** &    0.15*** &    0.13*** &    0.13*** &      0.14*** \\
\midrule                                               \multicolumn{ 7}{c}{\textbf{M6}} \\
\midrule
        \textbf{1m} &    0.01 &     -0.01 &     0.00 &     -0.01 &       0.00 &       0.02* \\

 \textbf{3m} &    0.10*** &    0.09*** &    0.07*** &    0.05*** &      0.04*** &       0.05** \\

 \textbf{6m} &    0.16*** &    0.13*** &    0.09*** &    0.09*** &    0.07** &      0.09** \\

\textbf{9m} &    0.21*** &    0.15*** &    0.11*** &    0.09*** &    0.09** &     0.11** \\

\textbf{12m} &    0.21*** &    0.19*** &    0.14*** &    0.11*** &     0.12*** &     0.14*** \\
\bottomrule
\end{tabular}
\end{center}
\noindent{\caption*{\footnotesize This table reports out-of-sample log predictive score (LS) across alternative models, at different prediction horizons, of $h$= 1-month, 3-month, 6-month, 9-month and 12-month. The seven forecasting models used are ATSM with alternative risk price restrictions. 
The EH implies the historical mean being the optimal forecast of excess returns. Positive values of this statistic imply that the forecast outperforms the historical mean forecast and suggests evidence of time-varying return predictability. Statistical significance is measured using a one-sided Diebold-Mariano statistic  computed with Newey-West standard errors. * denotes significance at 10\%, ** significance at 5\% and *** significance at 1\% level. The in-sample period is January 1990 to end of 2007, and the out-of-sample period starts in
January 2008 and ends at the end of 2018.}}
\end{table}
 
 Tables $\ref{table:LS2007}$ and $\ref{table:LS2018}$ report out-of-sample LS across models, maturities, and prediction horizons. Table $\ref{table:LS2007}$ summarises the results of the first sub-sample (i.e. 1985 - 2007), while Table $\ref{table:LS2018}$ reports results associated with the second sub-sample (i.e. 1990 - 2018). The main conclusions remain unchanged since results are qualitatively similar to those reported for the $R_{os}^2$ in the main body of the paper, showing evidence in favour of statistical predictability for all models regardless of the restrictions imposed. Results are more pronounced during the second sub-sample. In particular, in most cases, corresponding LS are positive and non-negligible, indicating statistical evidence of out-of-sample predictability. During the first sub-sample, the only model that performs poorly out-of-sample compared to the EH benchmark is the maximally flexible model, $M_0$, where LS are mostly negative, especially at longer maturities. This is not the case for the second sub-sample, where $M_0$ offers evidence of predictability, generating scores which are mostly positive, especially at the short end of the maturity spectrum and at longer prediction horizons. $M_1$ is performing very well in the second sub-sample but its superiority is challenged by $M_3$ in the first sub-sample. The sequential SSVS model that combines these two, $M_6$, does very well on both occasions although the differences between these models are quite small.

\newpage

\section*{Appendix E: Assessing Economic Performance - Additional Tests}

\subsection*{E.1: Investment Allocation Scenarios}

We investigate the economic evidence presented in the main body of the paper (see subsection 5.3) considering one more allocation scenario for investors. In this scenario, we follow \cite{Huang20} and restrict portfolio weights to the interval $[-1, 5]$ which sets maximum short-selling at $100\%$ while increasing the upper bound which now amounts to a maximum leveraging of $400\%$, thus relaxing the restriction associated with leveraging. 

Tables $\ref{table:CER2007c}$ and $\ref{table:CER2018c}$ present annualised CER values across models developed, at different maturities and investment horizons. In particular, Table $\ref{table:CER2007c}$ presents evidence under the first sub-sample (1985-2007), while Table $\ref{table:CER2018c}$ under the second sub-sample (1990-2018). Results remain qualitatively similar, yet more pronounced, compared to the main allocation scenario considered in the main body of the paper (i.e. w = [-1, 2]). In particular, results reveal that the maximally flexible model $M_0$ continues to fail offering any out-of-sample benefits to investors, generating CERs which are consistently negative across investment horizons and sub-periods. The situation is reversed for models with heavy restrictions on the dynamics of risk compensation (e.g. models $M_1$ and $M_2$) as well as for models inferred via the developed SSVS scheme (e.g. model $M_6$). Similar conclusions, yet more pronounced are revealed during the second sub-sample where economic benefits are substantially higher.
In all, results reveal that economic value can only be exploited if extreme and specific restrictions are imposed on the market price of risk specification. As such, results are robust to several allocation scenarios implemented.


\begin{table}[!htbp]\scriptsize						\caption{Out-of-sample Economic performance of Bond excess return forecasts across prediction horizons -
Investment scenario: w = [-1, 5] - Period: January 1985 - end of 2007.}
\label{table:CER2007c}
\begin{center}
\begin{tabular}{lllllll}
\toprule
\multicolumn{1}{m{1cm}}{\bf h} &\multicolumn{1}{m{0.5cm}}{\bf 2Y} &\multicolumn{1}{m{0.5cm}}{\bf 3Y} & \multicolumn{1}{m{0.5cm}}{\bf 4Y} & \multicolumn{1}{m{0.5cm}}{\bf 5Y} & \multicolumn{1}{m{0.5cm}}{\bf 7Y} & \multicolumn{ 1}{m{1cm}}{\bf 10Y}\\
\midrule                                             \multicolumn{7}{c}{$\mathbf{M_0}$}  \\
\midrule
\bf 1m &       0.05 &       0.22 &        0.43 &        -0.58 &         -1.96 &          -7.14 \\

\bf 3m &      -0.50 &      -0.25 &      -0.28 &      -1.00 &        -2.61 &          -6.42 \\

\bf 6m &      -0.80 &      -0.76 &      -0.92 &      -1.37 &       -2.74 &        -4.65 \\

\bf 9m &      -0.91 &      -0.79 &      -1.07 &      -1.45 &       -2.62 &         -3.84 \\

\bf 12m &         -0.58 &          -0.21 &         -0.57 &         -0.91 &         -2.32 &          -3.32 \\
\midrule
\multicolumn{7}{c}{$\mathbf{M_1}$}  \\
\midrule
\bf 1m &      0.10 &       0.23 &       0.50 &       0.94 &         1.20 &          1.05 \\

\bf 3m &     -0.12 &      0.35 &       0.80 &       1.05 &        1.02 &         0.82 \\

\bf 6m &      0.00 &      0.76 &      1.37* &      1.74** &      1.91** &       1.87** \\

\bf 9m &      -0.02 &      0.70* &      1.47** &     1.83*** &       2.15*** &       2.08*** \\

\bf 12m &         0.00 &         0.65** &         1.04* &         1.46** &         1.82*** &         1.81*** \\
\midrule
\multicolumn{7}{c}{$\mathbf{M_2}$} \\
\midrule
\bf 1m &        -0.04 &          0.43 &          -0.10 &          0.58 &          0.70 &          0.66 \\

\bf 3m &          -0.15 &         0.52 &          0.90 &          1.17 &          0.97 &         0.74 \\

\bf 6m &         -0.05 &         0.59 &        1.32** &       1.58** &       1.69** &       1.69** \\

\bf 9m &          -0.01 &         0.46* &        1.34** &       1.58*** &       1.87*** &      1.84*** \\

\bf 12m &        0.00 &        0.34* &        0.91* &         1.26** &     1.58*** &        1.58*** \\
\midrule
\multicolumn{7}{c}{$\mathbf{M_3}$} \\
\midrule
\bf 1m &      0.21 &         1.24 &        1.65 &         2.01 &          1.99 &          1.00 \\

\bf 3m &         0.12 &         0.95 &         1.31* &         1.44** &         1.02* &          0.24 \\

\bf 6m &        0.02 &         0.74* &          0.93** &        0.90** &       0.84** &      0.70** \\

\bf 9m &        0.00 &        0.59* &         0.93** &      0.76** &      0.79** &      0.68** \\

\bf 12m &       0.00 &        0.62** &        0.95** &        0.83* &          0.80** &      0.63 \\
\midrule
\multicolumn{7}{c}{$\mathbf{M_4}$} \\
\midrule
\bf 1m &          -0.21 &          0.10 &          0.18 &          0.34 &         0.35 &          -0.57 \\

\bf 3m &         0.01 &        0.49 &         1.07 &    1.24* &         1.02 &          0.22 \\

\bf 6m &        0.05 &         0.66 &          1.13* &         1.20* &      1.24 &      1.02 \\

\bf 9m &        -0.03 &         0.58 &          1.14** &       1.20** &      1.35* &      1.22** \\

\bf 12m &        -0.05 &        0.51* &        0.77 &         0.96** &        1.09* &      1.01** \\
\midrule
\multicolumn{7}{c}{$\mathbf{M_5}$} \\
\midrule
\bf 1m &         0.06 &       0.10 &         0.60 &          0.87 &          1.24 &          0.15 \\

\bf 3m &        -0.06 &        0.29 &          0.45 &     0.61 &          0.52 &          -0.23 \\

\bf 6m &         -0.06 &          0.39 &         0.57 &          0.67 &         0.71 &         0.58 \\

\bf 9m &        -0.01 &         0.43 &         0.61 &          0.62 &         0.74 &         0.71 \\

\bf 12m &        0.00 &        0.51 &          0.50 &        0.60 &          0.62 &         0.58 \\
\midrule
\multicolumn{7}{c}{$\mathbf{M_6}$} \\
\midrule
\bf 1m &      -0.06 &       0.53 &       0.80 &       1.33 &         1.42 &          1.22 \\

\bf 3m &      -0.02 &      0.64 &       0.98 &       1.28 &         1.11 &          0.78 \\

\bf 6m &      0.00 &      0.82* &      1.41** &      1.59** &       1.69** &         1.66** \\

\bf 9m &     0.00 &      0.72** &      1.50*** &      1.67*** &       1.91*** &        1.82*** \\

\bf 12m &         0.00 &         0.56*** &         1.03** &         1.32*** &         1.57*** &         1.52*** \\
\bottomrule
\end{tabular}
\end{center}
\noindent{\caption*{\scriptsize This table reports annualized certainty equivalent returns (CERs) across alternative models, at different prediction horizons, of $h$= 1-month, 3-month, 6-month, 9-month and 12-month. The coefficient of risk aversion is $\gamma=5$. CERs are generated by out-of-sample forecasts of bond excess returns and are reported in \%. At every time step, $t$, an investor with power utility preferences, evaluates the entire predictive density of bond excess returns and solves the asset allocation problem, thus optimally allocating her wealth between a riskless bond and risky bonds with maturities 2, 3, 4, 5, 7 and 10-years. CER is, then, defined as the value that equates the average utility of each alternative model against the average utility of the EH benchmark. The seven forecasting models used are ATSM with alternative risk price restrictions. Positive values indicate that the models perform better than the EH benchmark. In this scenario, portfolio weights are restricted to range in the interval [-1, 5], which amounts to maximum short-selling of 100\% and a maximum leveraging of 400\%. Statistical significance is measured using a one-sided Diebold-Mariano statistic  computed with Newey-West standard errors. * denotes significance at 10\%, ** significance at 5\% and *** significance at 1\% level. The in-sample period is January 1985 to end of 1996, and the out-of-sample period starts in January 1997 and ends at the end of 2007.}}
\end{table}

\begin{table}[!htbp]\scriptsize						\caption{Out-of-sample Economic performance of Bond excess return forecasts across prediction horizons - Investment scenario: w = [-1, 5] - Period: January 1990 - end of 2018.}
\label{table:CER2018c}
\begin{center}
\begin{tabular}{lllllll}
\toprule
\multicolumn{1}{m{1cm}}{\bf h} &\multicolumn{1}{m{0.5cm}}{\bf 2Y} &\multicolumn{1}{m{0.5cm}}{\bf 3Y} & \multicolumn{1}{m{0.5cm}}{\bf 4Y} & \multicolumn{1}{m{0.5cm}}{\bf 5Y} & \multicolumn{1}{m{0.5cm}}{\bf 7Y} & \multicolumn{ 1}{m{1cm}}{\bf 10Y}\\
\midrule                                             \multicolumn{7}{c}{$\mathbf{M_0}$}  \\
\midrule
\bf 1m &       -1.97 &       -2.79 &        -3.00 &        -2.88 &         -1.77 &          -5.23 \\

\bf 3m &      -0.56 &      -0.59 &      -0.46 &      -0.20 &        -0.46 &          -2.07 \\

\bf 6m &      0.02 &      -0.03 &      -0.08 &      0.18 &       0.26 &        -0.30 \\

\bf 9m &      0.13 &      -0.20 &      -0.46 &      -0.30 &       0.02 &         0.02 \\

\bf 12m &         0.20 &          0.18 &         -0.18 &         -0.06 &         -0.02 &          0.11 \\
\midrule
\multicolumn{7}{c}{$\mathbf{M_1}$}  \\
\midrule
\bf 1m &      -0.01 &       0.22 &       1.31 &       1.45 &         2.59 &          2.28 \\

\bf 3m &      0.00 &      0.63 &       1.68* &       2.41** &        2.78** &         3.41** \\

\bf 6m &      0.00 &      0.67** &      1.86*** &      2.42*** &      2.42*** &       3.02*** \\

\bf 9m &      0.00 &      0.43** &      2.00*** &      2.63*** &       2.65*** &       3.19*** \\

\bf 12m &         0.00 &         0.30*** &        1.79*** &         2.53*** &         2.45*** &         2.84*** \\
\midrule
\multicolumn{7}{c}{$\mathbf{M_2}$} \\
\midrule
\bf 1m &        0.06* &          0.31 &         1.02 &         1.29 &          2.54* &          1.70 \\

\bf 3m &          0.00 &          0.58 &         1.38 &          1.94* &          2.32** &         2.81* \\

\bf 6m &         0.00 &         0.59** &        1.50** &       1.92** &       1.90** &       2.48** \\

\bf 9m &          0.00 &         0.40** &        1.68*** &       2.13*** &       2.13*** &      2.64** \\

\bf 12m &        0.00 &        0.26*** &        1.57*** &         2.13*** &     2.02*** &        2.42*** \\
\midrule
\multicolumn{7}{c}{$\mathbf{M_3}$} \\
\midrule
\bf 1m &      -0.68 &         -3.12 &        -2.28 &         -2.03 &          0.83 &          -2.73 \\

\bf 3m &         -0.65 &         -2.26 &         -1.60 &         -1.17 &         -0.36 &          -2.70 \\

\bf 6m &        -0.36 &         -1.88 &          -1.78 &        -1.28 &       -0.63 &      -1.52 \\

\bf 9m &        -0.17 &        -1.61 &         -1.97 &      -1.75 &      -1.28 &      -1.55 \\

\bf 12m &       -0.06 &        -0.93 &       -1.52 &        -1.36 &         -1.32 &      -1.46 \\
\midrule
\multicolumn{7}{c}{$\mathbf{M_4}$} \\
\midrule
\bf 1m &          0.22 &          0.42 &          1.48 &          1.30 &        2.65* &          2.20 \\

\bf 3m &         0.28* &        0.83 &         1.84* &     2.36* &         2.65** &          3.00* \\

\bf 6m &        0.16* &         1.12** &          2.12*** &         2.58*** &      2.48** &      2.93** \\

\bf 9m &        0.08 &         0.93** &          2.25*** &       2.72*** &      2.68*** &      3.15*** \\

\bf 12m &        0.03 &        0.74** &        2.02*** &         2.58*** &        2.41*** &      2.80*** \\
\midrule
\multicolumn{7}{c}{$\mathbf{M_5}$} \\
\midrule
\bf 1m &         0.06 &       0.42 &         1.36 &          1.29 &          2.60* &          2.48 \\

\bf 3m &        0.02* &        0.70 &          1.77* &     2.35** &          2.69** &          3.09* \\

\bf 6m &         0.00 &          0.76** &         1.89*** &          2.37*** &         2.29** &        2.87*** \\

\bf 9m &        0.00 &         0.55** &         2.04*** &          2.57*** &         2.54*** &         3.00*** \\

\bf 12m &        0.00 &        0.39*** &          1.85*** &        2.53*** &          2.38*** &         2.72*** \\
\midrule
\multicolumn{7}{c}{\bf $\mathbf{M_6}$} \\
\midrule
\bf 1m &      0.04 &       0.32 &       1.87** &       1.74 &         2.99* &          2.46* \\

\bf 3m &      0.00 &      0.61 &       1.79** &       2.42** &         2.73** &          3.22* \\

\bf 6m &      0.00 &      0.61** &      1.77*** &      2.32*** &       2.28*** &         2.79*** \\

\bf 9m &      0.00 &      0.39** &      1.93*** &      2.50*** &       2.50*** &        2.99*** \\

\bf 12m &         0.00 &         0.28*** &         1.76*** &         2.44*** &         2.36*** &         2.77*** \\
\bottomrule
\end{tabular}
\end{center}
\noindent{\caption*{\scriptsize This table reports annualized certainty equivalent returns (CERs) across alternative models, at different prediction horizons, of $h$= 1-month, 3-month, 6-month, 9-month and 12-month. The coefficient of risk aversion is $\gamma=5$. CERs are generated by out-of-sample forecasts of bond excess returns and are reported in \%. At every time step, $t$, an investor with power utility preferences, evaluates the entire predictive density of bond excess returns and solves the asset allocation problem, thus optimally allocating her wealth between a riskless bond and risky bonds with maturities 2, 3, 4, 5, 7 and 10-years. CER is, then, defined as the value that equates the average utility of each alternative model against the average utility of the EH benchmark. The seven forecasting models used are ATSM with alternative risk price restrictions. Positive values indicate that the models perform better than the EH benchmark. In this scenario, portfolio weights are restricted to range in the interval [-1, 5], which amounts to maximum short-selling of 100\% and a maximum leveraging of 400\%. Statistical significance is measured using a one-sided Diebold-Mariano statistic  computed with Newey-West standard errors. * denotes significance at 10\%, ** significance at 5\% and *** significance at 1\% level. The in-sample period is January 1990 to end of 2007, and the out-of-sample period starts in January 2008 and ends at the end of 2018.}}
\end{table}

\newpage

\subsection*{E.2: Different Sets of Restrictions}

In this subsection we seek to assess the robustness of our results and the set of restrictions that have been inferred through the SSVS scheme developed. To that end we consider the possibility of using alternative benchmarks based on the market price of risk specifications in \cite{Cochrane09} and \cite{Duffee11}. These recent studies have investigated systematic approaches to imposing restrictions on the dynamics of risk compensation, in an attempt to improve upon the maximally flexible model. Hence, setting these models as benchmarks also allows comparisons with state-of-the-art approaches in the DTSM literature. The first model used has been suggested by \cite{Cochrane09} and is also one of the models with the highest posterior probability in \cite{Bauer18}. According to this model, only the elements $[\lambda_1]_{2,1}$ ($\lambda_{1,1}$ in our formulation) as well as the second element of vector $\lambda_{0}$ (the first element of vector $\lambda_{0\mathcal{P}}$ in our formulation), are left unrestricted, reflecting that risk premia are earned as compensation for exposure to level shocks only. The second set of restrictions we consider as benchmark, are the ones proposed in \cite{Duffee11}, allowing all parameters in the first row of matrix $\lambda_{1}$ ($\lambda$ in our formulation) as well as the first and second elements of vector $\lambda_{0}$ (of vector $\lambda_{0\mathcal{P}}$ in our formulation) to be unrestricted, suggesting that investors require compensation for level risk to all factors.

\begin{table}[!htbp]\scriptsize
\caption{Out-of-sample Economic performance of Bond excess return forecasts across prediction horizons 
}
\label{table:CERCPDUFFEE}
\begin{center}
\begin{tabular}{llllllllllllll}
\midrule \midrule 
\multicolumn{14}{c}{\bf Time Period: 1990 - 2018} \\
\midrule                                            
\multicolumn{7}{c}{\bf Panel 1: w= [-1,2]} & \multicolumn{7}{c}{\bf Panel 2: w= [-$\infty$,+$\infty$]} \\
\midrule
\multicolumn{1}{m{0.4cm}}{\bf h} &\multicolumn{1}{m{0.4cm}}{\bf 2Y} &\multicolumn{1}{m{0.4cm}}{\bf 3Y} & \multicolumn{1}{m{0.4cm}}{\bf 4Y} & \multicolumn{1}{m{0.4cm}}{\bf 5Y} & \multicolumn{1}{m{0.4cm}}{\bf 7Y} & \multicolumn{1}{m{0.4cm}}{\bf 10Y} & \multicolumn{1}{m{0.4cm}}{\bf h} &\multicolumn{1}{m{0.4cm}}{\bf 2Y} &\multicolumn{1}{m{0.4cm}}{\bf 3Y} & \multicolumn{1}{m{0.4cm}}{\bf 4Y} & \multicolumn{1}{m{0.4cm}}{\bf 5Y} & \multicolumn{1}{m{0.4cm}}{\bf 7Y} & \multicolumn{1}{m{0.4cm}}{\bf 10Y}\\
\midrule                                            
\multicolumn{14}{c}{\bf Restrictions as in Cochrane and Piazzesi (2009)}  \\
\midrule
\bf 1m &         -0.18 &         -0.62 &         -1.27 &         -1.96 &          -0.97 &          -3.77 &         \bf 1m &        -2.85 &          -3.95 &         -3.01 &         -2.88 &          -0.54 &          -3.93 \\

\bf 3m &         -0.04 &          -0.05 &          -0.31 &         -0.81 &          -0.66 &          -2.86 &         \bf 3m &          -1.55 &          -1.91 &         -1.37 &          -1.04 &         -0.45 &         -2.82 \\

\bf 6m &        -0.01 &         -0.06 &         -0.30 &        -0.68 &       -0.85 &        -1.58 &         \bf 6m &         -1.28 &         -1.69 &        -1.62 &       -1.21 &       -0.69 &       -1.58 \\

\bf 9m &          0.00 &          -0.06 &          -0.33 &          -0.74 &          -1.28 &        -1.51 &         \bf 9m &          -1.08 &         -1.59 &        -1.83 &       -1.64 &       -1.18 &      -1.50 \\

\bf 12m &         0.00 &         -0.03 &        -0.22 &         -0.57 &         -1.20 &          -1.34 &        \bf 12m &        -0.32 &        -0.80 &        -1.38 &         -1.23 &     -1.18 &        -1.33 \\
\midrule
\multicolumn{14}{c}{\bf Restrictions as in Duffee (2011)} \\
\midrule
\bf 1m &          -0.94 &          -1.44 &          -2.04 &          -2.73 &         -1.27 &         -2.67 &         \bf 1m &      -3.06 &         -2.68 &        -3.32 &         -3.41 &          -0.54 &          -1.68 \\

\bf 3m &          -0.37 &          -0.81 &         -1.37 &          -1.63 &          -1.10 &         -1.66 &         \bf 3m &         -0.75 &         -1.24 &         -1.81 &         -1.38 &         -0.53 &          -1.05 \\

\bf 6m &          -0.14 &          -0.49 &          -0.95 &          -1.06 &          -0.93 &        -0.62 &         \bf 6m &        0.49 &         -0.34 &          -1.14 &        -0.73 &       -0.41 &      -0.21 \\

\bf 9m &          -0.09 &          -0.36 &          -0.85 &          -0.93 &          -1.06 &          -0.41 &         \bf 9m &        0.61 &        -0.31 &         -1.21 &      -0.80 &      -0.52 &      -0.06 \\

\bf 12m &          0.02 &          -0.17 &         -0.66 &          -0.70 &      -0.85 &          -0.23 &        \bf 12m &       1.25 &        0.26 &        -0.96 &        -0.51 &         -0.39 &      0.11 \\
\midrule \midrule
\end{tabular}
\end{center}
\noindent{\caption*{\scriptsize This table reports annualized certainty equivalent returns (CERs), at different prediction horizons, of $h$= 1-month, 3-month, 6-month, 9-month and 12-month. The coefficient of risk aversion is $\gamma=5$. CERs are generated by out-of-sample forecasts of bond excess returns and are reported in \%. At every time step, $t$, an investor with power utility preferences, evaluates the entire predictive density of bond excess returns and solves the asset allocation problem, thus optimally allocating her wealth between a riskless bond and risky bonds with maturities 2, 3, 4, 5, 7 and 10-years. CER is, then, defined as the value that equates the average utility of each alternative model against the average utility of the EH benchmark. The two forecasting models used are the ATSM with sets of risk price restrictions as in \cite{Cochrane09} and \cite{Duffee11} respectively. Positive values indicate that the models perform better than the EH benchmark. Panel 1 presents CERs under the first scenario, where, portfolio weights are restricted to range in the interval [-1, 2], thus imposing maximum short-selling and leveraging of 100\% respectively, such that investors are prevented from extreme investments. Panel 2, reports CER values under the second scenario, where no allocation restrictions are imposed to investors and, as
such, portfolio weights are unbounded, thus allowing for maximum leveraging and short-selling. Statistical significance is measured using a one-sided Diebold-Mariano statistic  computed with Newey-West standard errors. * denotes significance at 10\%, ** significance at 5\% and *** significance at 1\% level. The in-sample period for results in Panel A is January 1985 to end of 1996, and the out-of-sample period starts in January 1997 and ends at the end of 2007. The in-sample period for results in Panel B is January 1990 to end of 2007, and the out-of-sample period starts in January 2008 and ends at the end of 2018.}}
\end{table}

Table $\ref{table:CERCPDUFFEE}$ reports results of annualised CERs for the two alternative market price of risk specifications. The coefficient of risk aversion is $\gamma = 5$, while portfolio weights are restricted to be in the interval $[-1, 2]$, thus imposing maximum short-selling and leveraging of 100\% respectively. There is clear evidence that neither of the two specifications/approaches are capable to offer any positive economic gains, out-of-sample, generating CER values that are consistently negative across the maturity spectrum and investment horizons. This is not the case, however, for DTSM models we have implemented which consistently deliver sizable economic gains especially at longer maturities. It seems more evident, that only with specific and extreme restrictions, as in $M_1$, it is possible to obtain some economic value from DTSMs.


\section*{Appendix F: Sequential Learning on a Predictive Regression Model}

Here we provide more details for the predictive regression model of Section 6, which can be written as
\begin{align}
    \label{VAR_b_app}
    \mathcal{P}_t - \mathcal{P}_{t-1} &= \mu + \Phi\mathcal{P}_{t-1} + \Sigma\varepsilon_t \\
rx_{t,t+h}^n &= a_h + \mathbf{b}_h'\mathcal{P}_{t+h-1}+\sigma_h\epsilon_t \label{eq:PR_app}
\end{align}
%
where $\mu$, $\Phi$ and $\Sigma$ are defined as before, $a_h$, $\sigma_h$ are scalars, and $\mathbf{b}_h$ is ($N\times1$) vector of the regression coefficients. For a given $h$ and information up to time $T$, the model defined by \eqref{VAR_b_app} and \eqref{eq:PR_app} can be estimated from the data $\{\mathcal{P}_t\}_{t=0}^T$, $\{rx_{t,t+h}^n\}_{t=0}^{T-h}$. Since the $\{\mathcal{P}_t\}_{t=0}^T$ are assumed to be directly observed, the overall likelihood is given by the product of VAR and predictive regression likelihoods obtained from \eqref{VAR_b_app} and \eqref{eq:PR_app} respectively. Hence we can view this model as running two independent models in parallel, a linear regression and a VAR model. 

We assign conjugate $g-$priors on the elements of $\mu$, $\Phi$, $a_h$ and $\mathbf{b}_h$, i.e. zero mean Normal priors with covariance matrices estimated being of the form $g\hat{V}$ where $\hat V$ is either $\sigma_h^2 (X^TX)^{-1}$ in the case of the predictive regression, with X denoting the relevant design matrix, whereas for the VAR model it is estimated as in the online appendix C.3 of \cite{Bauer18}. Following standard recommendations in the literature, $g$ is set to $\mbox{max}(T,p^2)$, where $p$ denotes the number of parameters. In our analyses, this translates to $T$, the length of the time series, in all cases. The choice of $g$-priors aims to guard against the Lindley paradox that affects the model evidences and therefore the posterior model probabilities. In order to impose sparsity we use the spike and slab priors on these parameters with slab variances being determined by the $g-$prior variances, whereas the spike variances are $10^4$ times smaller.

For the remaining parameters $\sigma_h^2$ and $\Sigma$ we use low informative inverse gamma $IG(0.01,0.01)$ and inverse Wishart $IW(\hat{\Sigma},N+1)$ distributions, where $\hat{\Sigma}$ is the MLE of $\Sigma$.

Given the priors above, it is possible to construct separate Gibbs samplers for each model, see online appendix C2 for the VAR case noting also that the full conditional of $\Sigma$ is another inverse Wishart distribution. A full sweep over both of these Gibbs samplers can then be used to sample from the posterior of the model defined by \eqref{VAR_b_app} and \eqref{eq:PR_app}. This Gibbs sampler can also be used for the jittering step of the IBIS algorithm operating on that model.

\end{appendices}

\end{document}